\documentclass[a4paper,10pt]{article}
\usepackage{amssymb}

\usepackage{amsmath}
\usepackage{fullpage}
\usepackage{boxedminipage}
\usepackage{listings}
\usepackage{minitoc}

\makeatletter \@addtoreset{equation}{section} \makeatother

\begin{document}

\begin{titlepage}

    \thispagestyle{empty}
    \begin{flushright}
        \hfill{CERN-PH-TH/2007-090} \\\hfill{UCLA/07/TEP/14}\\
    \end{flushright}

    \vspace{5pt}
    \begin{center}
        { \huge{\textbf{$\mathcal{N}=8$ non-BPS Attractors, Fixed Scalars\\\vspace{5pt}and Magic Supergravities}}}\vspace{25pt}
        \vspace{55pt}

        { \large{\textbf{{Sergio Ferrara$^{\diamondsuit\clubsuit\flat}$ and\ Alessio Marrani$^{\heartsuit\clubsuit\flat}$}}}}\vspace{15pt}

        {$\diamondsuit$ \it Physics Department,Theory Unit, CERN, \\
        CH 1211, Geneva 23, Switzerland\\
        \texttt{sergio.ferrara@cern.ch}}

        \vspace{15pt}

        {$\clubsuit$ \it INFN - Laboratori Nazionali di Frascati, \\
        Via Enrico Fermi 40,00044 Frascati, Italy\\
        \texttt{marrani@lnf.infn.it}}

        \vspace{10pt}

         {$\flat$ \it Department of Physics and Astronomy,\\
        University of California, Los Angeles, CA USA\\
        \texttt{ferrara@physics.ucla.edu}}

         \vspace{10pt}

        {$\heartsuit$ \it Museo Storico della Fisica e\\
        Centro Studi e Ricerche ``Enrico Fermi"\\
        Via Panisperna 89A, 00184 Roma, Italy}

        \vspace{50pt}
\end{center}


\begin{abstract}We analyze the Hessian matrix of the black hole
potential of  $\mathcal{N}=8$, $d=4$ supergravity, and determine its
rank at non-BPS critical points, relating the resulting spectrum to
non-BPS solutions (with non-vanishing central charge) of
$\mathcal{N}=2$, $d=4$ magic
supergravities and their ``mirror'' duals. We find agreement with the known degeneracy splitting of $%
\mathcal{N}=2$ non-BPS spectrum of generic special K\"{a}hler
geometries with cubic holomorphic prepotential. We also relate non-BPS critical points with vanishing central charge in $%
\mathcal{N}=2$ magic supergravities to a particular reduction of the $%
\mathcal{N}=8$, $\frac{1}{8}$-BPS critical points.
\end{abstract}

\end{titlepage}
\newpage\tableofcontents

\section{Introduction\label{Intro}\protect\smallskip}

After their discovery some time ago \cite{FKS}-\nocite{Strom,FK1,FK2}\cite
{FGK}, extremal black hole (BH) attractors have been object of intensive
study in the last years \cite{Sen-old1}-\nocite
{GIJT,Sen-old2,K1,TT,G,GJMT,Ebra1,K2,Ira1,Tom,BFM,AoB-book,FKlast,Ebra2,BFGM1,rotating-attr,K3,Misra1,Lust2,BFMY,CdWMa,DFT07-1,BFM-SIGRAV06,ADFT-2}
\cite{Saraikin-Vafa-1}. Such a flourishing development mainly can be
essentially traced back to new classes of solutions to the attractor
equations corresponding to non-BPS horizon geometries.

It has been recently realized that the ``effective black hole potential'' $%
V_{BH}$ of $\mathcal{N}\geqslant 2$-extended, $d=4$ supergravities exhibits
various species of critical points, whose supersymmetry-preserving and
stability features depend on the set of electric and magnetic BH charges.

For what concerns the case $\mathcal{N}=2$, critical points fall into three
distinct classes: ($\frac{1}{2}$-) BPS and two non-BPS classes, depending
whether the $\mathcal{N}=2$ central charge $Z$ vanishes or not at the BH
event horizon. The BPS critical points are known to be always stable (and
thus to give rise to actual attractor solutions), as far as they are points
at which the metric of the scalar manifold is positive-definite \cite{FGK}.

The stability not guaranteed in the non-BPS cases, in which the Hessian\ is
generally degenerate, \textit{i.e.} it exhibits some ``flat'' directions.
For example, for $\mathcal{N}=2$ supergravities whose vector multiplets'
scalar manifold is endowed with special K\"{a}hler (SK) $d$-geometries%
\footnote{%
Following the notation of \cite{dWVVP}, by $d$-geometry we mean a SK
geometry based on an holomorphic prepotential function of the cubic form $%
F\left( X\right) =d_{ABC}\frac{X^{A}X^{B}X^{C}}{X^{0}}$ ($A$, $B$, $%
C=0,1,...,n_{V}$).} of complex dimension $n_{V}$, it was shown in \cite{TT}
that the rank of the $2n_{V}\times 2n_{V}$ Hessian matrix of $V_{BH}$ (whose
real form is the scalar mass matrix) at the non-BPS $Z\neq 0$ critical
points has (at most) rank $n_{V}+1$ (corresponding to strictly positive
eigenvalues), with (at least) $n_{V}-1$ ``flat'' directions (\textit{i.e.}
vanishing eigenvalues).

Such a splitting ``$n_{V}+1$ / $n_{V}-1$'' of the non-BPS $Z\neq 0$ spectrum
has been confirmed in \cite{BFGM1}, where the $\mathcal{N}=2$ attractor
equations were studied in the framework of the homogeneous symmetric SK
geometries, which (apart from the case of the irreducible sequence based on
quadratic prepotential) are actually particular $d$-geometries.

In $\mathcal{N}>2$-extended, $d=4$ supergravities the BPS spectrum is
degenerate, too. As pointed out in \cite{ADF2}, the BPS splitting into
non-degenerate (with strictly positive eigenvalues) and ``flat'' (with
vanishing eigenvalues) directions can be explained respectively in terms of
the would-be vector multiplets' scalar and hypermultiplets' scalars of the $%
\mathcal{N}=2$ reduction of the considered $\mathcal{N}>2$ theory. For
example, in $\mathcal{N}=8$, $d=4$ supergravity (based on the coset manifold
$\frac{E_{7\left( 7\right) }}{SU\left( 8\right) }$) the $70\times 70$
Hessian of $V_{BH}$ at the (non-degenerate) $\frac{1}{8}$-BPS critical
points has rank $30$; its $30$ strictly positive and $40$ vanishing
eigenvalues respectively correspond to the $15$ vector multiplets and to the
$10$ hypermultiplets of the $\mathcal{N}=2$, $d=4$ spectrum obtained by
reducing $\mathcal{N}=8$ supergravity according to the following branching
of the $\mathbf{70}$ (four-fold antisymmetric) of $SU(8)$:
\begin{equation}
\begin{array}{l}
SU(8)\longrightarrow SU(6)\otimes SU(2); \\
\\
\mathbf{70}\longrightarrow \left( \mathbf{15},\mathbf{1}\right) \oplus
\left( \overline{\mathbf{15}},\mathbf{1}\right) \oplus \left( \mathbf{20},%
\mathbf{2}\right) ,
\end{array}
\label{SU(8)-->SU(2)xSU(6)}
\end{equation}
where $SU(6)\otimes SU(2)$ is nothing but the symmetry of the $8\times 8$ $%
\mathcal{N}=8$ central charge matrix $Z_{AB}$ (skew-diagonalizable in the
so-called ``normal frame'' \cite{Ferrara:1980ra}) at the considered
non-degenerate $\frac{1}{8}$-BPS critical points. $\mathbf{15}$, $\overline{%
\mathbf{15}}$ and $\mathbf{20}$ respectively are the two-fold antysimmetric,
its complex conjugate and the three-fold antysimmetric of $SU(6)$. In
general, the rank of the non-singular $\frac{1}{\mathcal{N}}$-BPS Hessian of
$V_{BH}$ in $2\leqslant \mathcal{N}\leqslant 8$-extended, $d=4$
supergravities is \cite{ADF2} $\left( \mathcal{N}-2\right) \left( \mathcal{N}%
-3\right) +2n_{V}$, where $n_{V}$ stands for the number of matter vector
multiplets (for $\mathcal{N}=6$, $n_{V}=1$ even though there are no vector
matter multiplets, because the extra singlet graviphoton counts as a matter
field).\medskip

The present paper is devoted to the study of the degeneracy of the non-BPS
Hessian of $V_{BH}$ in $\mathcal{N}=8$, $d=4$ supergravity, and of the
corresponding $\mathcal{N}=2$ theories obtained by consistent truncations.
Since such $\mathcal{N}=2$ theories content vector multiplets and
hypermultiplets which are some subsets of the kinematical reduction $%
\mathcal{N}=8\longrightarrow \mathcal{N}=2$ given by Eq. (\ref
{SU(8)-->SU(2)xSU(6)}), the massive and massless modes of the $\mathcal{N}=2$
non-BPS ($Z\neq 0$) Hessian must rearrange following the pattern of
degeneracy of the parent $\mathcal{N}=8$ supergravity, when reduced down to $%
\mathcal{N}=2$ theories.\medskip\

The plan of the paper is as follows.

In Sect. \ref{Sect2} we review the $\mathcal{N}=2$, $d=4$ \textit{magic}
models which can be obtained by consistent reduction of $\mathcal{N}=8$, $%
d=4 $ supergravity. Thence, Sect. \ref{Sect3} deals with the $\mathcal{N}=8$
(non-singular) $\frac{1}{8}$-BPS and non-BPS critical points of $V_{BH}$; in
particular, Subsect. \ref{Subsect3-1} reports known results on the $\mathcal{%
N}=8$, $d=4$ attractor equations and the(symmetries of the)ir solutions,
whereas Subsect. \ref{Subsect3-2} concerns the Hessian matrix of $V_{BH}$
both at $\frac{1}{8}$-BPS and non-BPS critical points. Thus, in Sect. \ref
{Sect4} we consider the $\mathcal{N}=2$ descendants of the $\mathcal{N}=8$, $%
\frac{1}{8}$-BPS critical points; they divide in $\mathcal{N}=2$, $\frac{1}{2%
}$-BPS and non-BPS $Z=0$ classes, whose spectra are both studied and
compared. In Sect. \ref{Sect5} we perform the same analysis for the
descendants of the $\mathcal{N}=8$ non-BPS critical points of $V_{BH}$,
\textit{i.e.} for the $\mathcal{N}=2$ non-BPS $Z\neq 0$ class of critical
points of $V_{BH,\mathcal{N}=2}$. We show that the interpretation of the
mass degeneracy splitting of $\mathcal{N}=8$ spectra in terms of $\mathcal{N}%
=2$ multiplets requires a different embedding of the $\mathcal{N}=2$ $%
\mathcal{R}$-symmetry $SU(2)_{H}$ in the $\mathcal{R}$-symmetry $SU(8)$ of
the parent $\mathcal{N}=8$ theory, depending on the structure and on the
eventual supersymmetry-breaking features of the considered class of
solutions to attractor equations. Our analysis also yields the
interpretation, in terms of the $U$-duality symmetry $E_{7(7)}$ of $\mathcal{%
N}=8$, $d=4$ supergravity, of the splitting ``$n_{V}+1$ / $n_{V}-1$'' of the
$2n_{V}$ eigenvalues of the $\mathcal{N}=2$ non-BPS $Z\neq 0$ Hessian matrix
for generic SK $d$-geometries of complex dimension $n_{V}$, found in \cite
{TT}. Finally, Sect. \ref{Conclusion} contains some general remarks, as well
as an outlook of possible future developments. \setcounter{equation}0

\section{$\mathcal{N}=8$ and $\mathcal{N}=2$ Magic Supergravities\label%
{Sect2}}

$\mathcal{N}=8$, $d=4$ supergravity is based on the $70$-dim. coset $\frac{G%
}{H}$, where the (continuous) $U$-duality group $G$ is $E_{7\left( 7\right)
} $ and its maximal compact subgroup (\textit{m.c.s.}) $H$ is $SU\left(
8\right) $, which is also the (local) $\mathcal{R}$-symmetry of the $%
\mathcal{N}=8$, $d=4$ supergravity. The vector and hyper multiplets' content
of an $\mathcal{N}=2$, $d=4$ reduction of $\mathcal{N}=8$, $d=4$
supergravity is given by a pair
\begin{equation}
\left( n_{V},n_{H}\right) \equiv \left( dim_{\mathbb{C}}\left( \frac{G_{V}}{%
H_{V}}\right) ,dim_{\mathbb{H}}\left( \frac{G_{H}}{H_{H}}\right) \right)
,~n_{V}\leqslant 15,~2n_{H}\leqslant 20,  \label{nV,nH}
\end{equation}
where $\frac{G_{V}}{H_{V}}$ and $\frac{G_{H}}{H_{H}}$ respectively stand for
the SK vector multiplets' scalar manifold and for the quaternionic
K\"{a}hler hypermultiplets' scalar manifold. Clearly, in order for the $%
\mathcal{N}=8\longrightarrow \mathcal{N}=2$ truncation to be consistent, the
isometry groups $G_{V}$ and $G_{H}$ of the two non-linear $\sigma $-models
should commute and should be both (proper) subgroups of $G=E_{7\left(
7\right) }$. We denote $H_{V}=m.c.s.\left( G_{V}\right) $ and $%
H_{H}=m.c.s.\left( G_{H}\right) $. Moreover, $H_{V}$ always contains a
factorized commuting $U(1)$ subgroup, which is promoted to global symmetry
(as the $G$s) when $n_{V}=0$; on the other hand, $H_{H}$ always contains a
factorized commuting $SU(2)$ subgroup, which is promoted to global symmetry
(as the $G$s) when $n_{H}=0$. As previously mentioned, $n_{V}=15$ and $%
n_{H}=10$ correspond to the reduction (\ref{SU(8)-->SU(2)xSU(6)}) of $%
\mathcal{N}=8$ supergravity, determining two $\mathcal{N}=2$ supergravities,
one based on $\frac{G_{V}}{H_{V}}=\frac{SO^{\ast }\left( 12\right) }{%
SU(6)\otimes U\left( 1\right) }$ with $\left( n_{V},n_{H}\right) =\left(
15,0\right) $, and the other one based on $\frac{G_{H}}{H_{H}}=\frac{%
E_{6\left( 2\right) }}{SU(6)\otimes SU\left( 2\right) }$ with $\left(
n_{V},n_{H}\right) =\left( 0,10\right) $.

In the following treatment we will consider only $\mathcal{N}=2$ \textit{%
maximal} supergravities, \textit{i.e.} $\mathcal{N}=2$ theories (obtained by
consistent truncations of $\mathcal{N}=8$ supergravity) which cannot be
obtained by a further reduction from some other $\mathcal{N}=2$ theory,
which are also \textit{magic}. They are called \textit{magic}, since their
symmetry groups are the groups of the famous \textit{Magic Square} of
Freudenthal, Rozenfeld and Tits associated with some remarkable geometries
\cite{Freudenthal2,magic}. From the analysis performed in \cite
{ADF,ADFFT,Ferrara-Gimon}, only six $\mathcal{N}=2$, $d=4$ \textit{maximal}
\textit{magic} supergravities\footnote{%
By $E_{7(p)}$ we denote a non-compact form of $E_{7}$, where $p\equiv \left(
\#\text{ non-compact}-\#\text{ compact}\right) $ generators of the group
\cite{Helgason,Gilmore}. In such a notation, the compact form of $E_{7}$ is $%
E_{7(-133)}$ ($dim_{\mathbb{R}}E_{7}=133$).} exist which can be obtained by
consistently truncating $\mathcal{N}=8$, $d=4$ supergravity; they are given%
\footnote{%
With a slight abuse of language we include among \textit{magic}
supergravities the $stu$ model, related to the Jordan algebra $\mathbb{R}%
\oplus \mathbf{\Gamma }_{2}=\mathbb{R}\oplus \mathbb{R}\oplus \mathbb{R}$,
which is the $n=0$ element of the sequence $\mathbb{R}\oplus \mathbf{\Gamma }%
_{2+n}$ of reducible Euclidean Jordan algebras of degree 3. $\mathbb{R}$
denotes the one dimensional Jordan algebra and $\mathbf{\Gamma }_{n+2}$
denotes the Jordan algebra of degree 2 associated with a quadratic form of
Lorentzian signature (see \textit{e.g.} Table 4 of \cite{BFGM1}, and Refs.
therein).
\par
Due to the group isomorphism $\frac{SO(2,2)}{SO(2)\otimes SO(2)}\sim \left(
\frac{SU(1,1)}{U(1)}\right) ^{2}$, the scalar manifold $\frac{G_{V}}{H_{V}}$
of the $stu$ model, corresponding to the element $n=0$ of the reducible SK
cubic sequence $\frac{SU(1,1)}{U(1)}\otimes \frac{SO(2,2+n)}{SO(2)\otimes
SO(2+n)}$ ($n\in \mathbb{N}\cup \left\{ 0,-1\right\} $, $dim_{\mathbb{C}%
}=n+3 $), is nothing but $\left( \frac{SU(1,1)}{U(1)}\right) ^{3}$.
\par
The image of $\left( \frac{SU(1,1)}{U(1)}\right) ^{3}$ through $c$-map is
given by the $4$-dim. (in $\mathbb{H}$) quaternionic manifold $\frac{SO(4,4)%
}{SO(4)\otimes SO(4)}$, which is the $\frac{G_{H}}{H_{H}}$ of the $stu$
model. Consistently, it is nothing but the element $n=0$ of the quaternionic
sequence $\frac{SO(4+n,4)}{SO(4+n)\otimes SO(4)}$ ($n\in \mathbb{N}\cup
\left\{ 0\right\} $, $dim_{\mathbb{H}}=n+1$), image of $\frac{SU(1,1)}{U(1)}%
\otimes \frac{SO(2,2+n)}{SO(2)\otimes SO(2+n)}$ through $c$-map (see \textit{%
e.g.} Table 4 of \cite{CFG}, and \cite{Sabharwal}).
\par
Finally, the $1$-dim. (in $\mathbb{H}$) quaternionic manifold $\frac{SU(2,1)%
}{SU(2)\otimes U(1)}$, corresponding to the $\frac{G_{H}}{H_{H}}$ of the
model $J_{3}^{\mathbb{H}}$, is the so-called \textit{universal hypermultiplet%
}, given by the $c$-map of the case $n_{V}=0$, \textit{i.e.} of \textit{pure
}$\mathcal{N}=2$, $d=4$ supergravity, which (among the homogeneous SK
geometries) is defined as the $\mathit{n=0}$\textit{\ limit} of the rank-$1$
sequence of quadratic irreducible SK manifolds $\frac{SU(1,n)}{U(1)\otimes
SU(n)}$ ($n\in \mathbb{N}$, $dim_{\mathbb{C}}=n$) \cite{Luciani}.} by Table
1.
\begin{table}[t]
\begin{center}
\begin{tabular}{|c||c|c|c|c|c|c|}
\hline
& $
\begin{array}{c}
\\
G_{V} \\
~
\end{array}
$ & $
\begin{array}{c}
\\
G_{H} \\
~
\end{array}
$ & $
\begin{array}{c}
\\
H_{V} \\
~
\end{array}
$ & $
\begin{array}{c}
\\
H_{H} \\
~
\end{array}
$ & $
\begin{array}{c}
~ \\
\frac{G_{V}}{H_{V}} \\
\otimes \\
\frac{G_{H}}{H_{H}} \\
~
\end{array}
$ & $
\begin{array}{c}
\\
\left( n_{V},n_{H}\right) \\
~
\end{array}
$ \\ \hline\hline
$J_{3}^{\mathbb{H}}$ & $
\begin{array}{c}
\\
SO^{\ast }(12) \\
~
\end{array}
$ & $SU(2)$ & $SU(6)\otimes U(1)$ & $-$ & $\frac{SO^{\ast }(12)}{%
SU(6)\otimes U(1)}$ & $\left( 15,0\right) $ \\ \hline
$J_{3}^{\mathbb{C}}$ & $SU(3,3)$ & $SU(2,1)$ & $
\begin{array}{c}
SU(3)\otimes SU(3) \\
\otimes \\
U(1)
\end{array}
$ & $SU(2)\otimes U(1)$ & $
\begin{array}{c}
~ \\
\frac{SU(3,3)}{SU(3)\otimes SU(3)\otimes U(1)} \\
\otimes \\
\frac{SU(2,1)}{SU(2)\otimes U(1)} \\
~
\end{array}
$ & $\left( 9,1\right) $ \\ \hline
$J_{3}^{\mathbb{R}}$ & $
\begin{array}{c}
\\
Sp\left( 6,\mathbb{R}\right) \\
~
\end{array}
$ & $G_{2(2)}$ & $SU(3)\otimes U(1)$ & $SU(2)\otimes SU(2)$ & $
\begin{array}{c}
~ \\
\frac{Sp(6,\mathbb{R})}{SU(3)\otimes U(1)} \\
\otimes \\
\frac{G_{2\left( 2\right) }}{SO(4)} \\
~
\end{array}
$ & $\left( 6,2\right) $ \\ \hline
$stu$ & $
\begin{array}{c}
SU(1,1) \\
\otimes \\
SO(2,2)
\end{array}
$ & $SO(4,4)$ & $
\begin{array}{c}
U(1) \\
\otimes \\
SO(2)\otimes SO(2)
\end{array}
$ & $SO(4)\otimes SO(4)$ & $
\begin{array}{c}
~ \\
\frac{SU(1,1)}{U(1)}\otimes \frac{SO(2,2)}{SO(2)\otimes SO(2)} \\
\otimes \\
\frac{SO(4,4)}{SO(4)\otimes SO(4)} \\
~
\end{array}
$ & $\left( 3,4\right) $ \\ \hline
$J_{3,M}^{\mathbb{R}}$ & $SU(1,1)$ & $F_{4(4)}$ & $U(1)$ & $USp(6)\otimes
SU(2)$ & $
\begin{array}{c}
~ \\
\frac{SU(1,1)}{U(1)} \\
\otimes \\
\frac{F_{4(4)}}{USp(6)\otimes SU(2)} \\
~
\end{array}
$ & $\left( 1,7\right) $ \\ \hline
$J_{3,M}^{\mathbb{C}}$ & $
\begin{array}{c}
\\
U(1) \\
~
\end{array}
$ & $E_{6(2)}$ & $-$ & $SU(6)\otimes SU(2)$ & $\frac{E_{6(2)}}{SU(6)\otimes
SU(2)}$ & $\left( 0,10\right) $ \\ \hline
\end{tabular}
\end{center}
\caption{\textbf{Data of the \textit{magic} }$\mathcal{N}=2$\textbf{,} $d=4$
\textbf{supergravities obtained as consistent truncation of (}$\frac{G}{H}=%
\frac{E_{7(7)}}{SU(8)}$\textbf{-based) }$\mathcal{N}=8$\textbf{,} $d=4$
\textbf{supergravity }}
\end{table}
The models have been denoted by referring to their SK geometry. $J_{3}^{%
\mathbb{H}}$, $J_{3}^{\mathbb{C}}$ and $J_{3}^{\mathbb{R}}$ stand for three
of the four $\mathcal{N}=2$, $d=4$ magic supergravities which, as their $5$%
-dim. versions, are respectively defined by the three simple Jordan algebras
$J_{3}^{\mathbb{H}}$, $J_{3}^{\mathbb{C}}$ and $J_{3}^{\mathbb{R}}$ of
degree 3 with irreducible norm forms, namely by the Jordan algebras of
Hermitian $3\times 3$ matrices over the division algebras of quaternions $%
\mathbb{H}$, complex numbers $\mathbb{C}$ and real numbers $\mathbb{R}$ \cite
{GST1,GST2,GST3,GST4,Jordan,Jacobson,Guna1,GPR}.

Since $E_{7(-25)}$ is a non-compact form of $E_{7}$ (as $E_{7(7)}$ is, as
well), the ``magic'' $\mathcal{N}=2$, $d=4$ supergravity defined by the
simple Jordan algebra $J_{3}^{\mathbb{O}}$ over the octonionic division
algebra $\mathbb{O}$, having vector multiplets' scalar manifold $\frac{%
E_{7(-25)}}{E_{6(-78)}\otimes SO(2)}$ ($dim_{\mathbb{C}}=27$), cannot be
obtained from $\mathcal{N}=8$, $d=4$ supergravity. Beside the analysis
performed in \cite{BFGM1}, Jordan algebras have been recently connected to
extremal black holes also in \cite{Rios}.

``$M$'' subscript denotes the model obtained by performing a $d=4$ \textit{%
mirror map} (\textit{i.e.} the composition of two $c$-maps in $d=4$) from
the original manifold; such an operation maps a model with content $\left(
n_{V},n_{H}\right) $ to a model with content $\left( n_{H}-1,n_{V}+1\right) $%
, and thus the mirror of $J_{3}^{\mathbb{H}}$, with $\left(
n_{V},n_{H}\right) =\left( -1,16\right) $ and quaternionic manifold $\frac{%
E_{7\left( -5\right) }}{SO(12)\otimes SU(2)}$ does not exist, \textit{at
least} in $d=4$. The $stu$ model \cite{Duff-stu,BKRSW,K3} is \textit{%
self-mirror}: $stu=stu_{M}$.

\section{\label{Sect3}$\mathcal{N}=8$, $d=4$ Critical Points and Hessian}

In Subsect. \ref{Subsect3-1} we will review the solutions to the attractor
equations of $\mathcal{N}=8$, $d=4$ supergravity, mainly following \cite
{FKlast} (see \cite{ADFT} for a recent review of Attractor Mechanism in $%
\mathcal{N}\geqslant 2$-extended, $d=4$ supergravities). Thence, in Subsect.
\ref{Subsect3-2}we will consider the related critical spectrum given by the
Hessian of $V_{BH}$; while the non-singular $\frac{1}{8}$-case was
investigated in \cite{ADF2} (see also \cite{ADF3}), the non-BPS case was
hitherto unknown.

\subsection{\label{Subsect3-1}Solutions to Attractor Equations}

The black hole potential of $\mathcal{N}=8$, $d=4$ supergravity (based on
the real coset $\frac{E_{7\left( 7\right) }}{SU\left( 8\right) }$) \cite
{Cremmer:1979up} reads as follows \cite{Andrianopoli:1996ve,FGK} ($%
A,B=1,\dots ,8$ throughout):
\begin{equation}
V_{BH}=\frac{1}{2}Z_{AB}\overline{Z}^{AB},  \label{N8pot}
\end{equation}
where $Z_{AB}$ (and its complex conjugate $\overline{Z}^{AB}$) is the
central charge matrix (and its conjugate), sitting in the two-fold
antisymmetric complex $\mathbf{28}$ of $E_{7(7)}$. It depends on $70\left(
=dim_{\mathbb{R}}\left( \frac{E_{7\left( 7\right) }}{SU\left( 8\right) }%
\right) \right) $ real scalars $\phi ^{i}$ ($i=1,\dots ,70$ throughout,
unless otherwise noted), where the local $SU(8)$ symmetry was used to remove
63 scalars from the representation $\mathbf{133}$ of scalars in $E_{7(7)}$.

The $SU(8)$-covariant derivatives \cite{Andrianopoli:1996ve} of the central
charge matrix are defined by the \textit{Maurer-Cartan equations} for $\frac{%
E_{7\left( 7\right) }}{SU\left( 8\right) }$:
\begin{equation}
D_{i}Z_{AB}=\frac{1}{2}\overline{Z}^{CD}P_{ABCD,i}\Leftrightarrow D_{i}%
\overline{Z}^{AB}=\frac{1}{2}Z_{CD}\overline{P}_{,i}^{ABCD},
\label{derivative}
\end{equation}
where $P_{ABCD}=P_{i,[ABCD]}d\phi ^{i}$ is the $70\times 70$ vielbein 1-form
of ${\frac{E_{7(7)}}{SU(8)}}$, sitting in the $\mathbf{70}$ (four-fold
antisymmetric) of the stabylizer $SU(8)$, and satisfying to the \textit{%
self-dual reality} condition
\begin{equation}
\overline{P}^{ABCD}=\frac{1}{4!}\epsilon ^{ABCDEFGH}P_{EFGH}\Leftrightarrow
P_{ABCD}=\frac{1}{4!}\epsilon _{ABCDEFGH}\overline{P}^{EFGH},  \label{self}
\end{equation}
$\epsilon _{ABCDEFGH}$ being the rank-$8$ completely antisymmetric
Ricci-Levi-Civita tensor of $SU(8)$. By using Eqs. (\ref{derivative}) and (%
\ref{self}), and by exploiting the invertibility (non-singularity) of $%
P_{ABCD,i}$, the criticality conditions for $V_{BH}$ can be rewritten as
\cite{Andrianopoli:1996ve,FGK,FKlast}
\begin{equation}
\overline{Z}^{[AB}\overline{Z}^{CD]}+{\frac{1}{4!}}\epsilon
^{ABCDEFGH}Z_{[EF}Z_{GH]}=0,  \label{critical2}
\end{equation}
which are usually referred to as the $\mathcal{N}=8$, $d=4$ attractor
equations. They are purely algebraic in the $\left( Z_{AB},\overline{Z}%
^{AB}\right) $, and they hold for all non-singular (\textit{i.e.} with $%
V_{BH}\neq 0$) critical points of $V_{BH}$ in $\frac{E_{7\left( 7\right) }}{%
SU\left( 8\right) }$ at which $P_{ABCD,i}$ is invertible.

The local $SU(8)$ symmetry allows one to go to the so-called ``normal
frame'' \cite{Ferrara:1980ra}. In such a frame, $Z_{AB}$ and the unique
Cartan-Cremmer-Julia quartic invariant $J_{4}$ \cite{Cartan, Cremmer:1979up}
of the fundamental representation $\mathbf{56}$ of $E_{7\left( 7\right) }$
respectively read as follows ($\epsilon \equiv \left(
\begin{array}{cc}
0 & -1 \\
1 & 0
\end{array}
\right) $ is the $2$-dim. symplectic metric):
\begin{equation}
\begin{array}{l}
Z_{AB,normal}=
\begin{pmatrix}
z_{1}\epsilon & 0 & 0 & 0 \\
0 & z_{2}\epsilon & 0 & 0 \\
0 & 0 & z_{3}\epsilon & 0 \\
0 & 0 & 0 & z_{4}\epsilon
\end{pmatrix}
\equiv
\begin{pmatrix}
\rho _{1}\epsilon & 0 & 0 & 0 \\
0 & \rho _{2}\epsilon & 0 & 0 \\
0 & 0 & \rho _{3}\epsilon & 0 \\
0 & 0 & 0 & \rho _{4}\epsilon
\end{pmatrix}
e^{i\varphi /4}; \\
\\
z_{i}\equiv \rho _{i}e^{i\varphi /4}\in \mathbb{C},~\rho _{i}\in \mathbb{R}%
^{+},~i=1,2,3,4, \\
\\
\rho _{1}\geqslant \rho _{2}\geqslant \rho _{3}\geqslant \rho _{4}\geqslant
0,~\varphi \in \left[ 0,8\pi \right) .
\end{array}
~  \label{ZAB}
\end{equation}
\begin{equation}
J_{4,normal}=\Big [(\rho _{1}+\rho _{2})^{2}-(\rho _{3}+\rho _{4})^{2}\Big]%
\Big [(\rho _{1}-\rho _{2})^{2}-(\rho _{3}-\rho _{4})^{2}\Big]+8\rho
_{1}\rho _{2}\rho _{3}\rho _{4}(cos\varphi -1).
\end{equation}
Note that $Z_{AB,normal}$ has an $\left( SU(2)\right) ^{4}$ symmetry. The $%
\mathcal{N}=8$ attractor equations (\ref{critical2}) acquire the following
simple form \cite{FKlast}:
\begin{equation}
\left\{
\begin{array}{l}
z_{1}z_{2}+\overline{z_{3}}\overline{z_{4}}=0; \\
z_{1}z_{3}+\overline{z_{2}}\overline{z_{4}}=0; \\
z_{2}z_{3}+\overline{z_{1}}\overline{z_{4}}=0.
\end{array}
\right.  \label{attractors}
\end{equation}
As expected from the analysis of \cite{FM,FG}, $\mathcal{N}=8$, $d=4$
extremal black hole attractor equations (\ref{attractors}) have only 2
distinct classes of non-singular solutions ($\frac{1}{8}$-BPS for $J_{4}>0$,
non-BPS for $J_{4}<0$):\medskip

\textbf{1. }$\frac{1}{8}$\textbf{-BPS:}
\begin{equation}
\rho _{1}=\rho _{\frac{1}{8}-BPS}\in \mathbb{R}_{0}^{+},~\varphi _{\frac{1}{8%
}-BPS}\in \left[ 0,8\pi \right) ,~\rho _{2,\frac{1}{8}-BPS}=\rho _{3,\frac{1%
}{8}-BPS}=\rho _{4,\frac{1}{8}-BPS}=0.  \label{primera-1}
\end{equation}
The corresponding orbit of supporting BH charges in the $\mathbf{56}$ of $%
E_{7(7)}$ is $\mathcal{O}_{\frac{1}{8}-BPS}=\frac{E_{7(7)}}{E_{6(2)}}$, with
$J_{4,normal,\frac{1}{8}-BPS}=\rho _{\frac{1}{8}-BPS}^{4}>0$ and classical
entropy $S_{BH,\frac{1}{8}-BPS}=\pi \sqrt{J_{4,normal,\frac{1}{8}-BPS}}=\pi
\rho _{\frac{1}{8}-BPS}^{2}$. As implied by Eq. (\ref{primera-1}), $%
Z_{AB,normal,\frac{1}{8}-BPS\text{ }}$ has symmetry enhancement $\left(
SU(2)\right) ^{4}\longrightarrow SU(6)\otimes SU(2)=m.c.s.\left(
E_{6(2)}\right) $.\smallskip\ Notice that $\varphi _{\frac{1}{8}-BPS}$ is
actually undetermined.

\textbf{2. non-BPS: }
\begin{equation}
\rho _{1,non-BPS}=\rho _{2,non-BPS}=\rho _{3,non-BPS}=\rho _{4,non-BPS}=\rho
_{non-BPS}\in \mathbb{R}_{0}^{+},\text{ \ }\varphi _{non-BPS}=\pi .
\label{segunda-1}
\end{equation}
The corresponding orbit of supporting BH charges in the $\mathbf{56}$ of $%
E_{7(7)}$ is $\mathcal{O}_{non-BPS}=\frac{E_{7(7)}}{E_{6(6)}}$, with $%
J_{4,normal,non-BPS}=-16\rho _{non-BPS}^{4}<0$ and classical entropy $%
S_{BH,non-BPS}=\pi \sqrt{-J_{4,normal,non-BPS}}=4\pi \rho _{non-BPS}^{2}$.
The deep meaning of the extra factor $4$ in $S_{BH,non-BPS}$ as compared to $%
S_{BH,\frac{1}{8}-BPS}$ can be clearly explained when considering the
so-called ``$stu$ interpretation'' of $\mathcal{N}=8$ regular critical
points \cite{FKlast}. As implied by Eq. (\ref{segunda-1}), $%
Z_{AB,normal,non-BPS\text{ }}$ has symmetry enhancement $\left( SU(2)\right)
^{4}\longrightarrow USp(8)=m.c.s.\left( E_{6(6)}\right) $; indeed
\begin{equation}
Z_{AB,normal,non-BPS\text{ }}=e^{i\frac{\pi }{4}}\rho _{non-BPS}\Omega _{AB},
\label{segunda-2}
\end{equation}
where $\Omega _{AB}$ is the $USp(8)$ metric:
\begin{equation}
\Omega _{AB}\equiv \left(
\begin{array}{cccc}
\epsilon &  &  &  \\
& \epsilon &  &  \\
&  & \epsilon &  \\
&  &  & \epsilon
\end{array}
\right) .
\end{equation}

Thus, as pointed out at the end of the Introduction of \cite{BFGM1}, the
symmetry of $Z_{AB,normal\text{ }}$ gets enhanced at the particular points
of $\frac{E_{7(7)}}{SU(8)}$ given by the non-singular solutions of $\mathcal{%
N}=8$, $d=4$ attractor equations (\ref{attractors}). In general, \textit{the
invariance properties of the non-singular solutions to attractor eqs. are
given by the m.c.s. of the stabilizer of the corresponding supporting BH
charge orbit}.

\subsection{\label{Subsect3-2}Critical Spectra}

Let us now consider the Hessian of $V_{BH}$. By further covariantly
differentiating $V_{BH}$, one gets \cite{ADF2}
\begin{equation}
H_{ij}\equiv D_{i}D_{j}V_{BH}=\frac{1}{2}Z_{CD}\overline{Z}^{AB}\overline{P}%
_{,j}^{CDEF}P_{ABEF,i}=H_{ji}.  \label{N=8-Hessian}
\end{equation}
\smallskip

\textbf{1. }$\frac{1}{8}$\textbf{-BPS:}

By recalling Eq. (\ref{primera-1}), it can be computed that ($a,b=3,...,8$)
\cite{ADF2}
\begin{equation}
\begin{array}{l}
H_{ij,\frac{1}{8}-BPS}=\frac{1}{2}\left[ Z_{CD}\overline{Z}^{AB}\overline{P}%
_{,j}^{CDEF}P_{ABEF,i}\right] _{\frac{1}{8}-BPS}= \\
\\
=2\rho _{\frac{1}{8}-BPS}^{2}\left[ \overline{P}_{,j}^{12ab}P_{12ab,i}\right]
_{\frac{1}{8}-BPS}=\frac{1}{12}\rho _{\frac{1}{8}-BPS}^{2}\epsilon
^{12abEFGH}\left[ P_{EFGH,j}P_{12ab,i}\right] _{\frac{1}{8}-BPS}.
\end{array}
\label{N=8-Hessian-BPS}
\end{equation}
As observed in \cite{ADF2}, the pattern of degeneracy of the modes of $H_{ij,%
\frac{1}{8}-BPS}$ can be understood by noticing that the very structure of
the non-singular $\frac{1}{8}$-BPS solution (\ref{primera-1}), in which only
one eigenvalue of the skew-diagonal matrix $Z_{AB,normal}$ is not vanishing,
yields that the $\mathcal{N}=8$ theory effectively reduces to an $\mathcal{N}%
=2$ theory. Consequently, the degeneracy splitting of the eigenvalues of $%
H_{ij,\frac{1}{8}-BPS}$ will respect the multiplicity of the $\mathcal{N}=2$
scalar degrees of freedom: the ``flat'' directions will correspond to the $%
\mathcal{N}=2$ hypermultiplet content, whereas the ``non-flat'' directions
(with strictly positive eigenvalues) will correspond to the $\mathcal{N}=2$
vector multiplet content.

The crucial point is the choice of the kinematical reduction $\mathcal{N}%
=8\longrightarrow \mathcal{N}=2$. As previously mentioned, in the
non-singular $\frac{1}{8}$-BPS case it is performed through the branching of
$\mathbf{70}$ of $SU(8)$ along the $\frac{1}{8}$-BPS enhanced symmetry $%
SU(6)\otimes SU(2)$ given by Eq. (\ref{SU(8)-->SU(2)xSU(6)}), yielding:

\textit{i}) $2n_{V}=30$ strictly positive directions (\textit{massive
Hessian modes}), corresponding to $15$ complex $\mathcal{N}=2$ vector
multiplets' scalars, sitting into the $\left( \mathbf{15},\mathbf{1}\right)
\oplus \left( \overline{\mathbf{15}},\mathbf{1}\right) $ of $SU(6)\otimes
SU(2)$, and parameterized by the $30$ real components $P_{abcd}$;

and

\textit{ii}) $4n_{H}=40$ ``flat'' directions (\textit{massless Hessian modes}%
), corresponding to $10$ quaternionic $\mathcal{N}=2$ hypermultiplets'
scalars, sitting into the $\left( \mathbf{20},\mathbf{2}\right) $ of $%
SU(6)\otimes SU(2)$, and parameterized by the $40$ real components\footnote{%
Notice that, due to the self-dual reality condition (\ref{self}), $P_{12ab}$
can be re-expressed in terms of the other independent component of $P_{ABCD}$%
.} $\left\{ P_{1abc},P_{2abc}\right\} $.

Thus, at $\mathcal{N}=2$, $\frac{1}{2}$-BPS critical points of $V_{BH,%
\mathcal{N}=2}$ Eq. (\ref{SU(8)-->SU(2)xSU(6)}) can be written as follows:
\begin{equation}
\mathbf{70}\longrightarrow \underset{\text{vectors' scalars}}{\overset{m\neq
0}{\overbrace{\left( \mathbf{15},\mathbf{1}\right) }}\oplus \overset{m\neq 0%
}{\overbrace{\left( \overline{\mathbf{15}},\mathbf{1}\right) }}}\oplus
\overset{m=0}{\overbrace{\underset{\text{hypers' scalars}}{\left( \mathbf{20}%
,\mathbf{2}\right) }}},  \label{BPS-mass-splitting}
\end{equation}

Under the branching (\ref{SU(8)-->SU(2)xSU(6)}) $P_{ABCD}$ decomposes as $%
P_{ABCD}\longrightarrow \left\{ P_{1abc},P_{2abc},P_{abcd}\right\} $. As it
holds true in general (also at non-BPS non-singular critical points), the $%
\mathcal{N}=2$ vector and hyper scalar degrees of freedom are respectively
singlets and doublets of the $\mathcal{N}=2$ $\mathcal{R}$-symmetry $SU(2)_{%
\mathcal{R},\mathcal{N}=2}\equiv SU(2)_{H}$, which in general lies inside
the whole $\mathcal{N}=8$ $\mathcal{R}$-symmetry $SU(8)$.

Thus, in the non-singular $\mathcal{N}=8$, $\frac{1}{8}$-BPS case \textit{all%
} $\mathcal{N}=2$ vector multiplets' scalar degrees of freedom of $H_{ij}$
are massive, while \textit{all} its $\mathcal{N}=2$ hypermultiplets' scalar
degrees of freedom are massless; this can be understood by observing that
the preservation of 4 supersymmetric degrees of freedom forces such two
different kind of $\mathcal{N}=2$ degrees of freedom to follow separated
mass degeneracy patterns.\smallskip

\textbf{2. non-BPS:}

The same can be intuitively guessed \textit{not }to hold in the
(non-singular) non-BPS case, where no supersymmetric degrees of freedom are
preserved by the critical solution. In fact, what actually happens is that,
for what concerns the mass degeneracy spliiting, the $\mathcal{N}=2$ vector
and hyper scalar degrees of freedom of $H_{ij}$ mix together, in a way which
follows the various possibilities yielded by \textit{all} the \textit{maximal%
} \textit{magic} $\mathcal{N}=2$, $d=4$ supergravities which are consistent
truncations of $\mathcal{N}=8$, $d=4$ supergravity (given by Table 1).

Indeed, by recalling Eqs. (\ref{segunda-1}) and (\ref{segunda-2}), it can be
computed that
\begin{equation}
\begin{array}{l}
H_{ij,non-BPS}=\frac{1}{2}\left[ Z_{CD}\overline{Z}^{AB}\overline{P}%
_{,j}^{CDEF}P_{ABEF,i}\right] _{non-BPS}= \\
\\
=\frac{1}{2}\rho _{non-BPS}^{2}\left[
\begin{array}{l}
\frac{4}{27}\epsilon ^{ABCDEFGH}P_{\left[ ABCD\right| ,i}P_{\left| EFGH%
\right] ,j}+ \\
\\
+\left( 32-\frac{1}{18}\right) P_{ABCD,i}P_{EFGH,j}\Omega ^{\lbrack
AB}\Omega ^{CD]}\Omega ^{\lbrack EF}\Omega ^{GH]}
\end{array}
\right] _{non-BPS}.
\end{array}
\label{N=8-Hessian-non-BPS}
\end{equation}

In this case, the relevant branching of the $\mathbf{70}$ of the stabylizer $%
SU(8)$ is along the non-BPS enhanced symmetry $USp(8)$:
\begin{equation}
\begin{array}{l}
SU(8)\longrightarrow USp(8); \\
\\
\mathbf{70}\longrightarrow \mathbf{42}\oplus \mathbf{27}\oplus \mathbf{1},
\end{array}
\label{SU(8)-->USp(8)}
\end{equation}
where $\mathbf{42}$, $\mathbf{27}$ and $\mathbf{1}$ respectively are the
four-fold antysimmetric (traceless), two-fold antysimmetric (traceless) and
the singlet of $USp(8)$. Under the branching (\ref{SU(8)-->USp(8)}) $%
P_{ABCD} $ decomposes as follows:
\begin{equation}
\begin{array}{l}
P_{ABCD}\longrightarrow \left\{ \hat{P}_{ABCD},\hat{P}_{AB},\hat{P}%
^{0}\right\} ; \\
\\
\left\{
\begin{array}{l}
\mathbf{1}\text{~of }USp(8):\hat{P}^{0}\equiv \frac{1}{2^{4}}P_{ABCD}\Omega
^{\lbrack AB}\Omega ^{CD]}; \\
\mathbf{27}\text{~of }USp(8):\hat{P}_{AB}\equiv \frac{3}{2}P_{ABCD}\Omega
^{CD}-3\hat{P}^{0}\Omega _{AB},~\hat{P}_{AB}=\hat{P}_{\left[ AB\right] },~%
\hat{P}_{AB}\Omega ^{AB}=0; \\
\mathbf{42}\text{~of }USp(8):\hat{P}_{ABCD}\equiv P_{ABCD}-\hat{P}%
_{[AB}\Omega _{CD]}-\hat{P}^{0}\Omega _{\lbrack AB}\Omega _{CD]},~\hat{P}%
_{ABCD}=\hat{P}_{\left[ ABCD\right] },~\hat{P}_{ABCD}\Omega ^{CD}=0.
\end{array}
\right.
\end{array}
\label{P-USp(8)-decomp}
\end{equation}
By using such an $USp(8)$-covariant decomposition of $P_{ABCD}$, the result (%
\ref{N=8-Hessian-non-BPS}) can be rewritten as follows:
\begin{equation}
H_{ij,non-BPS}=\frac{1}{2}\rho _{non-BPS}^{2}\left[ \left( \frac{2}{3}%
\right) ^{4}\overline{\hat{P}}_{,j}^{AB}\hat{P}_{AB,i}+2^{13}\hat{P}_{,i}^{0}%
\hat{P}_{,j}^{0}\right] _{non-BPS},  \label{N=8-Hessian-non-BPS-2}
\end{equation}
where the barred quantities have definitions and properties analogue to the
ones in Eq. (\ref{P-USp(8)-decomp}), to which they are related by the
self-dual reality condition (\ref{self}), too.

Thus, one sees that the non-BPS kinematical reduction $\mathcal{N}%
=8\longrightarrow \mathcal{N}=2$ performed through the branching of $\mathbf{%
70}$ of $SU(8)$ along the non-BPS enhanced symmetry$USp(8)$ given by Eq. ((%
\ref{SU(8)-->USp(8)})) yields a different mass degeneracy splitting with
respect to the $\frac{1}{8}$-BPS case treated above. Indeed, as evident from
Eq. (\ref{N=8-Hessian-non-BPS-2}), $H_{ij,non-BPS}$ is splitted in:

\textit{i}) $28$ strictly positive directions (\textit{massive Hessian modes}%
), sitting into the $\mathbf{27}\oplus \mathbf{1}$ of $USp(8)$, and
parameterized by the $27+1$ real components $\hat{P}_{AB}$ and $\hat{P}^{0}$;

and

\textit{ii}) $42$ ``flat'' directions (\textit{massless Hessian modes}),
sitting into the $\mathbf{42}$ of $USp(8)$, and parameterized by the $42$
real components $\hat{P}_{ABCD}$.

Thus, at $\mathcal{N}=8$ non-BPS critical points of $V_{BH}$ Eq. (\ref
{SU(8)-->USp(8)}) can be written as follows:
\begin{equation}
\mathbf{70}\longrightarrow \overset{m=0}{\overbrace{\mathbf{42}}}\oplus
\overset{m\neq 0}{\overbrace{\mathbf{27}}}\oplus \overset{m\neq 0}{%
\overbrace{\mathbf{1}}}.  \label{non-BPS-Z<>0-mass-splitting}
\end{equation}
As we will see below, the identification of the massive and massless Hessian
modes with the $\mathcal{N}=2$ vector multiplets' and hypermultiplets'
scalars is model-dependent.

However, from the splitting ``$n_{V}+1$ / $n_{V}-1$'' found in \cite{TT}
(holding for generic SK $d$-geometries), we can state the following result
for non-BPS $Z\neq 0$ critical points of all $\mathcal{N}=2$, $d=4$
supergravities listed in Table 1: given a pair $\left( n_{V},n_{H}\right) $
describing the multiplets' content of the model, $4n_{H}+n_{V}-1$ massless
real modes sit in the $\mathbf{42}$ of $USp(8)$, while $n_{V}$ real massive
modes sit in the $\mathbf{27}$ of $USp(8)$ (the remaining $1$ real massive
mode sitting in the singlet $\mathbf{1}$ of $USp(8)$).

\section{\label{Sect4}\protect\smallskip $\mathcal{N}=8$, $\frac{1}{8}$-BPS
Critical Points and their $\mathcal{N}=2$ Descendants}

As pointed out above, $\mathcal{N}=8$, $\frac{1}{8}$-BPS critical points of $%
V_{BH}$ have symmetry $SU(6)\otimes SU(2)_{\mathcal{R}}$, where $SU(2)_{%
\mathcal{R}}$ is the $SU(2)$ factor of the $\mathcal{N}=8$ $\mathcal{R}$%
-symmetry $SU(8)$ which commutes with $SU(6)$. The $70\times 70$ $\frac{1}{8}
$-BPS Hessian matrix $H_{ij,\frac{1}{8}-BPS}$ of $V_{BH}$ has rank $30$,
corresponding to the $\mathcal{N}=8\longrightarrow \mathcal{N}=2$
kinematical decomposition (\ref{SU(8)-->SU(2)xSU(6)}). It is worth noticing
that, under the same branching, the $\mathbf{56}$ fundamental representation
of the $\mathcal{N}=8$ $U$-duality group $G=E_{7(7)}$ decomposes into
representation of the $\frac{1}{8}$-BPS symmetry $SU(6)\otimes SU(2)_{%
\mathcal{R}}$ as follows:
\begin{equation}
\mathbf{56}\longrightarrow \left( \mathbf{15},\mathbf{1}\right) \oplus
\left( \overline{\mathbf{15}},\mathbf{1}\right) \oplus \left( \mathbf{1},%
\mathbf{1}\right) \oplus \left( \mathbf{1},\mathbf{1}\right) \oplus \left(
\mathbf{6},\mathbf{2}\right) \oplus \left( \overline{\mathbf{6}},\mathbf{2}%
\right) ,  \label{56}
\end{equation}
which consistently gives $16$ electric and $16$ magnetic charges for the $%
15+1$ Abelian vectors of the $\mathcal{N}=2$ matter and gravity
supermultiplets. The remaining charges from the decomposition (\ref{56})
pertain to the graviphotons which are partners of the $6$ remaining
gravitino multiplets $6\left( \frac{3}{2},2\left( 1\right) ,\frac{1}{2}%
\right) $ in the $\mathcal{N}=8\longrightarrow \mathcal{N}=2$ reduction (\ref
{SU(8)-->SU(2)xSU(6)}), which precisely have $\left( \mathbf{6},\mathbf{2}%
\right) \oplus \left( \overline{\mathbf{6}},\mathbf{2}\right) $ electric and
magnetic field strenghts.

\subsection{\label{Subsect4-1}$\mathcal{N}=2$, $\frac{1}{2}$-BPS}

For the $\mathcal{N}=2$, $d=4$ supergravities listed in Table 1, the
enhanced symmetry $\mathcal{S}_{\frac{1}{2}-BPS}$ of $\mathcal{N}=2$, $d=4$ $%
\frac{1}{2}$-BPS critical points of $V_{BH,\mathcal{N}=2}$ is given by \cite
{ADF2,BFGM1}\label{jazz}
\begin{equation}
\mathcal{S}_{\frac{1}{2}-BPS}=H_{0}\otimes H_{H},  \label{N=2-BPS-symmetry}
\end{equation}
where $H_{0}$ is the stabylizer of the $\mathcal{N}=2$, $\frac{1}{2}$%
-BPS-supporting BH charge orbit\footnote{%
Here and in the following treatment we will make use of the notation set up
in \cite{BFGM1}. $H_{0}$ is defined (for $n_{V}\neq 0$) as $H_{0}\equiv
\frac{H_{V}}{U(1)}$ \cite{BFGM1}.}, and $H_{H}$ is the stabylizer of $\frac{%
G_{H}}{H_{H}}$. Furthermore, $\mathcal{N}=2$, $\frac{1}{2}$-BPS case has $%
\mathcal{N}=2$ quartic $G_{V}$-invariant $I_{4}>0$, where $I_{4}$ is nothing
but a suitable ``truncation'' of the $E_{7(7)}$-invariant $J_{4}$. Since the
sign of the $U$-duality group invariant (built out from the symplectic
representation of the $U$-duality group) does not change in the $\mathcal{N}%
=8\longrightarrow \mathcal{N}=2$ supersymmetry reduction, it is clear that
the $\mathcal{N}=2$, $\frac{1}{2}$-BPS case comes from the reduction of the $%
\mathcal{N}=8$, $\frac{1}{8}$-BPS case.

Thus, $\mathcal{S}_{\frac{1}{2}-BPS}$ must be included in the overall
enhanced symmetry $SU(6)\otimes SU(2)_{\mathcal{R}}$ of the $\mathcal{N}=8$,
$\frac{1}{8}$-BPS case:
\begin{equation}
\mathcal{S}_{\frac{1}{2}-BPS}\subseteq SU(6)\otimes SU(2)_{\mathcal{R}}.
\label{N=2-BPS-symmetry-incl}
\end{equation}
The very structure of the quaternionic K\"{a}hler manifold $\frac{G_{H}}{%
H_{H}}$ yields that $H_{H}$ always include at least one explicit factor $%
SU(2)$, which is promoted to a global symmetry in the case $n_{H}=0$. Thus, $%
H_{H}$ can always (for $n_{H}\neq 0$) be rewritten as
\begin{equation}
H_{H}=\frac{H_{H}}{SU(2)}\otimes SU(2).  \label{HH}
\end{equation}
In general, the $\mathcal{N}=2$ $\mathcal{R}$-symmetry group $SU(2)_{%
\mathcal{R},\mathcal{N}=2}$ is identified with the $SU(2)$ factorized in the
r.h.s. of Eq. (\ref{HH}), which in the follow we will denote with the
subscript ``$H$'':
\begin{equation}
SU(2)_{\mathcal{R},\mathcal{N}=2}=SU(2)_{H}\subseteq H_{H}.
\end{equation}
The identification determining the $\mathcal{N}=2$, $\frac{1}{2}$-BPS case
as descendant of the $\mathcal{N}=8$, $\frac{1}{8}$-BPS case reads as
follows (recall Eq. (\ref{primera-1})):
\begin{equation}
Z_{12,\frac{1}{8}-BPS}\equiv z_{1,\frac{1}{8}-BPS}=e^{i\varphi /4}\rho _{%
\frac{1}{8}-BPS}=Z_{\frac{1}{2}-BPS}\in \mathbb{C}_{0}.
\end{equation}
Therefore, at $\mathcal{N}=2$, $\frac{1}{2}$-BPS critical points of $V_{BH,%
\mathcal{N}=2}$ (which preserve 4 supersymmetry charges, and are always
stable \cite{FGK}, thus corresponding to attractor configurations), the $%
\mathcal{N}=8\longrightarrow \mathcal{N}=2$ kinematical decomposition (\ref
{SU(8)-->SU(2)xSU(6)}) identifies $SU(2)_{\mathcal{R}}$ on the r.h.s. of Eq.
(\ref{N=2-BPS-symmetry-incl}) with the $\mathcal{N}=2$ $\mathcal{R}$%
-symmetry $SU(2)_{H}$:
\begin{equation}
SU(2)_{\mathcal{R}}=SU(2)_{H}.
\end{equation}

Thus, Eq. (\ref{HH}) can be rewritten as
\begin{equation}
H_{H}=\frac{H_{H}}{SU(2)_{\mathcal{R}}}\otimes SU(2)_{\mathcal{R}},
\end{equation}
which, by Eq. (\ref{N=2-BPS-symmetry-incl}), implies that
\begin{equation}
H_{0}\otimes \frac{H_{H}}{SU(2)_{\mathcal{R}}}\subseteq SU(6).
\end{equation}
The corresponding data for all the $\mathcal{N}=2$, $d=4$ supergravities
which are consistent truncations of the $\mathcal{N}=8$, $d=4$ theory
(listed in Table 1) are given in Table 2 (for the columns ``$\mathcal{O}_{%
\frac{1}{2}-BPS}$'' and ``$H_{0}$'' refer to Tables 3 and 8 of \cite{BFGM1}).

\begin{table}[t]
\begin{center}
\begin{tabular}{|c||c|c|c|}
\hline
& $
\begin{array}{c}
\\
\frac{1}{2}\text{-BPS orbit } \\
~~\mathcal{O}_{\frac{1}{2}-BPS}=\frac{G_{V}}{H_{0}} \\
~
\end{array}
$ & $
\begin{array}{c}
\\
H_{0}\equiv \frac{H_{V}}{U(1)} \\
~
\end{array}
$ & $
\begin{array}{c}
\\
\frac{H_{H}}{SU(2)_{\mathcal{R}}=SU(2)_{H}} \\
~
\end{array}
$ \\ \hline\hline
$
\begin{array}{c}
\\
J_{3}^{\mathbb{H}} \\
~
\end{array}
$ & $\frac{SO^{\ast }(12)}{SU(6)}~$ & $
\begin{array}{c}
\\
SU(6) \\
~
\end{array}
$ & $\nexists H_{H},~~SU_{H}(2)=SU(2)_{\mathcal{R}}=G_{H}~$ \\ \hline
$
\begin{array}{c}
\\
J_{3}^{\mathbb{C}} \\
~
\end{array}
$ & $\frac{SU(3,3)}{SU(3)\otimes SU(3)}~$ & $
\begin{array}{c}
\\
SU(3)\otimes SU(3) \\
~
\end{array}
~$ & $U(1)$ \\ \hline
$
\begin{array}{c}
\\
J_{3}^{\mathbb{R}} \\
~
\end{array}
$ & $\frac{Sp(6,\mathbb{R})}{SU(3)}$ & $
\begin{array}{c}
\\
SU(3) \\
~
\end{array}
$ & $SU(2)$ \\ \hline
$
\begin{array}{c}
\\
stu \\
~
\end{array}
$ & $\frac{\left( SU(1,1)\right) ^{3}}{\left( U(1)\right) ^{2}}~$ & $
\begin{array}{c}
\\
\left( U(1)\right) ^{2} \\
~
\end{array}
$ & $\left( SU(2)\right) ^{3}$ \\ \hline
$
\begin{array}{c}
\\
J_{3,M}^{\mathbb{R}} \\
~
\end{array}
$ & ${SU(1,1)}{}$ & $\mathbb{I}$ & $
\begin{array}{c}
\\
USp(6) \\
~
\end{array}
~$ \\ \hline
$
\begin{array}{c}
\\
J_{3,M}^{\mathbb{C}} \\
~
\end{array}
$ & $-$ & $-$ & $
\begin{array}{c}
\\
SU(6) \\
~
\end{array}
$ \\ \hline
\end{tabular}
\end{center}
\caption{\textbf{The }$\frac{1}{2}$\textbf{-BPS supporting BH charge orbit} $%
\mathcal{O}_{\frac{1}{2}-BPS}$\textbf{, and the compact groups }$H_{0}$
\textbf{and }$\frac{H_{H}}{SU(2)_{\mathcal{R}}}$ \textbf{(relevant at }$%
\mathcal{N}=2$\textbf{,} $\frac{1}{2}$\textbf{-BPS critical points) for the}
$\mathcal{N}=2$\textbf{,} $d=4$ \textbf{supergravities listed in Table 1}}
\end{table}

From Table 2 it is also evident that $SU(2)_{\mathcal{R}}$ has necessarly to
be chosen in $H_{H}$, because in all models $H_{0}$ does not contain a
factorized $SU(2)$. Moreover, two orders of considerations follow:

\textit{i}) $H_{0}\otimes \frac{H_{H}}{SU(2)_{\mathcal{R}}}$ is a \textit{%
proper} subgroup of $SU(6)$ in all models but the two limit models $J_{3}^{%
\mathbb{H}}$ (having $n_{H}=0$, and thus $H_{H}$ undefined) and $J_{3,M}^{%
\mathbb{C}}$ (having $n_{V}=0$, and thus $H_{0}$ undefined and corresponding
to a Reissner-N\"{o}rdstrom extremal BH, only having $\frac{1}{2}$-BPS
critical points).

For $J_{3}^{\mathbb{H}}$, $SU(2)_{\mathcal{R}}=SU(2)_{H}$ is identified with
the global symmetry $SU(2)=G_{H}$ due to $n_{H}=0$.

On the other hand, for $J_{3,M}^{\mathbb{C}}$ it holds that $\mathcal{S}_{%
\frac{1}{2}-BPS}=H_{H}=SU(6)\otimes SU(2)_{\mathcal{R}}$, \textit{i.e.} the
enhanced $\mathcal{N}=2$, $\frac{1}{2}$-BPS symmetry $\mathcal{S}_{\frac{1}{2%
}-BPS}$, the stabylizer of the quaternionic K\"{a}hler manifold $\frac{G_{H}%
}{H_{H}}$ and the enhanced $\mathcal{N}=8$, $\frac{1}{8}$-BPS symmetry
coincide.

\textit{ii}) Two models exist where an \textit{apriori} arbitrariness in the
identification of $SU(2)_{H}$ in $H_{H}$ exists: $J_{3}^{\mathbb{R}}$ and $%
stu$.

However, in $J_{3}^{\mathbb{R}}$ such an arbitrariness is removed by the
quantum numbers of the hypermultiplets' scalars (which are always doublets
of $SU(2)_{H}$); the ``right'' $SU(2)$ to choose is the one promoted to a
global symmetry in the limit case $n_{H}=0$. On the other hand, in $stu$
case the arbitrariness of choice is removed by the noteworthy \textit{%
triality symmetry} of the model.

\subsection{\label{Subsect4-2}$\mathcal{N}=2$ non-BPS $Z=0$}

For the $\mathcal{N}=2$, $d=4$ supergravities listed in Table 1, the overall
symmetry $\mathcal{S}_{non-BPS,Z=0}$ of $\mathcal{N}=2$, $d=4$ non-BPS $Z=0$
critical points of $V_{BH,\mathcal{N}=2}$ is given by \cite{BFGM1}
\begin{equation}
\mathcal{S}_{non-BPS,Z=0}=\widetilde{h}^{\prime }\otimes H_{H},
\end{equation}
where $\widetilde{h}^{\prime }$ is the \textit{m.c.s.} (factorized by $U(1)$%
) of the stabylizer $\widetilde{H}$ of the $\mathcal{N}=2$ non-BPS $Z=0$%
-supporting BH charge orbit \cite{BFGM1}. Furthermore, $\mathcal{N}=2$
non-BPS $Z=0$ case has $\mathcal{N}=2$ quartic $G_{V}$-invariant $I_{4}>0$,
as the $\mathcal{N}=2$, $\frac{1}{2}$-BPS case. Thus, it is clear that $%
\mathcal{N}=2$ non-BPS $Z=0$ case comes from the very same $\mathcal{N}%
=8\longrightarrow \mathcal{N}=2$ supersymmetry reduction giving raise to $%
\mathcal{N}=2$, $\frac{1}{2}$-BPS case. Thus, $\mathcal{S}_{non-BPS,Z=0}$
must be included in the overall enhanced symmetry $SU(6)\otimes SU(2)_{%
\mathcal{R}}$ of the $\mathcal{N}=8$, $\frac{1}{8}$-BPS case:
\begin{equation}
\mathcal{S}_{non-BPS,Z=0}\subseteq SU(6)\otimes SU(2)_{\mathcal{R}}.
\label{N=2-non-BPS-Z=0-symmetry-incl}
\end{equation}
The identification determining the $\mathcal{N}=2$ non-BPS $Z=0$ case as
descendant of the $\mathcal{N}=8$, $\frac{1}{8}$-BPS case reads as follows
(recall that $Z_{non-BPS,Z=0}=0$):
\begin{equation}
Z_{12,\frac{1}{8}-BPS}\equiv z_{1,\frac{1}{8}-BPS}=e^{i\varphi /4}\rho _{%
\frac{1}{8}-BPS}=\left( D_{i}Z\right) _{non-BPS,Z=0}\neq 0,
\end{equation}
where $i$ is one particular element of the set $\left\{ 1,...,n_{V}\right\} $%
. In this sense, the key difference with respect to the previously treated $%
\mathcal{N}=2$, $\frac{1}{2}$-BPS case is that the $\mathcal{N}=2$ central
charge is interchanged with one $\mathcal{N}=2$ \textit{matter charge}.

This leads to the fact that for $\mathcal{N}=2$ models under consideration
which exhibit ``flat'' Hessian directions at $\mathcal{N}=2$ non-BPS $Z=0$
critical points of $V_{BH,\mathcal{N}=2}$ (namely $J_{3}^{\mathbb{H}}$, $%
J_{3}^{\mathbb{C}}$ and $J_{3}^{\mathbb{R}}$) the $SU(2)_{\mathcal{R}}$ of
the enhanced $\mathcal{N}=8$, $\frac{1}{8}$-BPS symmetry $SU(2)_{\mathcal{R}%
}\otimes SU(6)$ is not identified with the $SU(2)_{\mathcal{R},\mathcal{N}%
=2} $ (\textit{i.e.} with (one of) the $SU(2)$(s) factorized in $H_{H}$) any
more, but rather it is identified with an explicit $SU(2)$ factor in $%
\widetilde{h}^{\prime }$. Thus, for these models $\widetilde{h}^{\prime }$
can be rewritten as
\begin{equation}
J_{3}^{\mathbb{H}},J_{3}^{\mathbb{C}},J_{3}^{\mathbb{R}}:\widetilde{h}%
^{\prime }=\frac{\widetilde{h}^{\prime }}{SU(2)}\otimes SU(2).
\label{factor-non-BPS-Z=0}
\end{equation}
By making the identification $SU(2)_{\mathcal{R}}=SU(2)$ factor on the
r.h.s. of Eq. (\ref{factor-non-BPS-Z=0}), one can thus rewrite Eq. (\ref
{N=2-non-BPS-Z=0-symmetry-incl}) as follows:
\begin{equation}
J_{3}^{\mathbb{H}},J_{3}^{\mathbb{C}},J_{3}^{\mathbb{R}}:\frac{\widetilde{h}%
^{\prime }}{SU(2)_{\mathcal{R}}}\otimes H_{H}\subseteq SU(6).
\label{factor-non-BPS-Z=0-2}
\end{equation}

For what concerns the remaining models, $J_{3,M}^{\mathbb{C}}$ and $J_{3,M}^{%
\mathbb{R}}$ respectively have $n_{V}=0,1$ and thus they do not have $%
\mathcal{N}=2$ non-BPS $Z=0$ critical points of $V_{BH,\mathcal{N}=2}$ at
all.

The $stu$ model has $\widetilde{h}^{\prime }=SO(2)$, and thus Eqs. (\ref
{factor-non-BPS-Z=0}) and (\ref{factor-non-BPS-Z=0-2}) do not hold. In such
a model all goes the same way as for the previously treated $\mathcal{N}=2$,
$\frac{1}{2}$-BPS case, and consequently in $stu$ model $\mathcal{N}=2$
non-BPS $Z=0$ critical points of $V_{BH,\mathcal{N}=2}$ are stable, \textit{%
i.e.} there are no ``flat'' non-BPS $Z=0$ Hessian directions at all. This
can be simply understood by noticing that in such an $\mathcal{N}=2$
framework \textit{triality symmetry} puts non-BPS $Z=0$ critical points on
the very same footing of $\frac{1}{2}$-BPS critical points, which are always
stable and thus do not have any ``flat'' direction at all.

The corresponding data for all the \textit{maximal magic} $\mathcal{N}=2$, $%
d=4$ supergravities which are consistent truncations of the $\mathcal{N}=8$,
$d=4$ theory (listed in Table 1) are given in Table 3 (for the column ``$%
\widetilde{h}^{\prime }$'' refer to Table 8 of \cite{BFGM1}).
\begin{table}[t]
\begin{center}
\begin{tabular}{|c||c|c|c|}
\hline
& $
\begin{array}{c}
\\
\text{non-BPS }Z=0\text{ orbit} \\
\mathcal{O}_{non-BPS,Z=0}=\frac{G_{V}}{\widetilde{H}}~ \\
~
\end{array}
$ & $
\begin{array}{c}
\\
\widetilde{h}^{\prime }\equiv \frac{m.c.s.\left( \widetilde{H}\right) }{U(1)}
\\
~
\end{array}
$ & $
\begin{array}{c}
\\
\frac{H_{H}}{SU(2)_{H}} \\
~
\end{array}
$ \\ \hline\hline
$
\begin{array}{c}
\\
J_{3}^{\mathbb{H}} \\
~
\end{array}
$ & $\frac{SO^{\ast }(12)}{SU(4,2)}$ & $
\begin{array}{c}
\\
SU(4)\otimes SU(2)_{\mathcal{R}} \\
~
\end{array}
$ & $\nexists H_{H},~~SU_{H}(2)=G_{H}~$ \\ \hline
$
\begin{array}{c}
\\
J_{3}^{\mathbb{C}} \\
~
\end{array}
$ & $\frac{SU(3,3)}{SU(2,1)\otimes SU(1,2)}~$ & $
\begin{array}{c}
\\
SU(2)\otimes SU(2)_{\mathcal{R}}\otimes U(1) \\
~
\end{array}
~$ & $U(1)$ \\ \hline
$
\begin{array}{c}
\\
J_{3}^{\mathbb{R}} \\
~
\end{array}
$ & $\frac{Sp(6,\mathbb{R})}{SU(2,1)}$ & $
\begin{array}{c}
\\
SU(2)_{\mathcal{R}} \\
~
\end{array}
$ & $SU(2)~$ \\ \hline
$
\begin{array}{c}
\\
stu \\
~
\end{array}
$ & $\frac{\left( SU(1,1)\right) ^{3}}{\left( U(1)\right) ^{2}}$ & $
\begin{array}{c}
\\
SO(2) \\
~
\end{array}
~$ & $\left( SU(2)\right) ^{2}\otimes SU(2)_{\mathcal{R}}~$ \\ \hline
$
\begin{array}{c}
\\
J_{3,M}^{\mathbb{R}} \\
~
\end{array}
$ & $-$ & $-$ & $
\begin{array}{c}
\\
USp(6), \\
~ \\
SU(2)_{H}=SU(2)_{\mathcal{R}}
\end{array}
~$ \\ \hline
$
\begin{array}{c}
\\
J_{3,M}^{\mathbb{C}} \\
~
\end{array}
$ & $-$ & $-$ & $
\begin{array}{c}
\\
SU(6), \\
\\
SU(2)_{H}=SU(2)_{\mathcal{R}}
\end{array}
$ \\ \hline
\end{tabular}
\end{center}
\caption{\textbf{The non-BPS }$Z=0$ \textbf{supporting BH charge orbit }$%
\mathcal{O}_{non-BPS,Z=0}$\textbf{, and the} \textbf{compact groups }$%
\widetilde{h}^{\prime }$ \textbf{and }$\frac{H_{H}}{SU(2)_{H}}$ \textbf{%
(relevant at }$\mathcal{N}=2$\textbf{\ non-BPS }$Z=0$ \textbf{critical
points) for the} $\mathcal{N}=2$\textbf{,} $d=4$ \textbf{supergravities
listed in Table 1}}
\end{table}

Let us consider two explicit examples, namely the models $J_{3}^{\mathbb{H}}$
and $stu$.\medskip

The model $J_{3}^{\mathbb{H}}$ has the highest number of vector multiplets ($%
n_{V}=15$) and no hypermultiplets at all ($n_{H}=0$); thus, $H_{H}$ cannot
be defined, and $SU(2)=SU(2)_{H}$ is promoted to a global symmetry, which
here coincides with $G_{H}$ itself. $SU(2)_{\mathcal{R}}$ is identified with
the factor $SU(2)$ in $\widetilde{h}^{\prime }=SU(4)\otimes SU(2)$, thus it
holds that $SU(4)\otimes G_{H}=SU(4)\otimes SU(2)_{H}\subset SU(6)$. The $%
\mathbf{15}$, $\overline{\mathbf{15}}$ and $\mathbf{20}$ of $SU(6)$
decompose under $SU(4)\otimes SU(2)_{H}$ as follows:
\begin{equation}
\begin{array}{l}
\mathbf{15}=\left( \mathbf{4},\mathbf{2}\right) \oplus \left( \mathbf{6},%
\mathbf{1}\right) \oplus \left( \mathbf{1},\mathbf{1}\right) ; \\
\\
\overline{\mathbf{15}}=\left( \overline{\mathbf{4}},\mathbf{2}\right) \oplus
\left( \mathbf{6},\mathbf{1}\right) \oplus \left( \mathbf{1},\mathbf{1}%
\right) ; \\
\\
\mathbf{20}=\left( \mathbf{4},\mathbf{1}\right) \oplus \left( \overline{%
\mathbf{4}},\mathbf{1}\right) \oplus \left( \mathbf{6},\mathbf{2}\right) .
\end{array}
\end{equation}

Thus, by also recalling Eq. (\ref{BPS-mass-splitting}), one obtains that at $%
\mathcal{N}=2$ non-BPS $Z=0$ critical points the $\mathcal{N}=8$, $\frac{1}{8%
}$-BPS enhanced symmetry $SU(6)\otimes SU(2)_{\mathcal{R}}$ decomposes under
$SU(4)\otimes SU(2)_{H}\otimes SU(2)_{\mathcal{R}}$ as follows:
\begin{equation}
\begin{array}{l}
m\neq 0:\left( \mathbf{15},\mathbf{1}\right) \oplus \left( \overline{\mathbf{%
15}},\mathbf{1}\right) =\left( \mathbf{4},\mathbf{2},\mathbf{1}\right)
\oplus \left( \overline{\mathbf{4}},\mathbf{2},\mathbf{1}\right) \oplus
\left( \mathbf{6},\mathbf{1},\mathbf{1}\right) \oplus \left( \mathbf{6},%
\mathbf{1},\mathbf{1}\right) \oplus \left( \mathbf{1},\mathbf{1},\mathbf{1}%
\right) \oplus \left( \mathbf{1},\mathbf{1},\mathbf{1}\right) ; \\
\\
m=0:\left( \mathbf{20},\mathbf{2}\right) =\left( \mathbf{4},\mathbf{1},%
\mathbf{2}\right) \oplus \left( \overline{\mathbf{4}},\mathbf{1},\mathbf{2}%
\right) \oplus \left( \mathbf{6},\mathbf{2},\mathbf{2}\right) .
\end{array}
\label{non-BPS-Z=0-masses}
\end{equation}

As previously mentioned, in general the $\mathcal{N}=2$ vector multiplets'
and hypermultiplets' scalar degrees of freedom are respectively given by the
singlets and doublets of $SU(2)_{H}$. For the model under consideration, all
vector multiplets' scalars are included in the $\mathcal{N}=2$, $d=4$
spectrum, whereas all hypermultiplets' scalars are truncated away by
dimensional reduction $\mathcal{N}=8\longrightarrow \mathcal{N}=2$. Thus,
the representation decomposition (\ref{non-BPS-Z=0-masses}) yields that at $%
\mathcal{N}=2$ non-BPS $Z=0$ critical points the vector multiplets' scalars
and hypermultiplets' scalars respectively sit in the following
representations of $SU(4)\otimes SU(2)_{H}\otimes SU(2)_{\mathcal{R}}$:
\begin{eqnarray}
&&
\begin{array}{l}
\underset{\text{(all in the }\mathcal{N}=2\text{, }d=4\text{ spectrum) }}{30%
\text{ (real) vectors' scalar degrees of freedom~}}=~\left\{
\begin{array}{l}
\overset{14~~m\neq 0}{~\overbrace{\left( \mathbf{6},\mathbf{1},\mathbf{1}%
\right) \oplus \left( \mathbf{6},\mathbf{1},\mathbf{1}\right) \oplus \left(
\mathbf{1},\mathbf{1},\mathbf{1}\right) \oplus \left( \mathbf{1},\mathbf{1},%
\mathbf{1}\right) }}\oplus \\
\\
\oplus \overset{16~~m=0}{\overbrace{\left( \mathbf{4},\mathbf{1},\mathbf{2}%
\right) \oplus \left( \overline{\mathbf{4}},\mathbf{1},\mathbf{2}\right) }};
\end{array}
\right. \\
\\
\underset{\text{(all truncated away in the }\mathcal{N}=8\longrightarrow
\mathcal{N}=2\text{ reduction)}}{40\text{ (real) hypers' scalar degrees of
freedom}}~=~\overset{16~~m\neq 0}{\overbrace{\left( \mathbf{4},\mathbf{2},%
\mathbf{1}\right) \oplus \left( \overline{\mathbf{4}},\mathbf{2},\mathbf{1}%
\right) }}\oplus \overset{24~~m=0}{\overbrace{\left( \mathbf{6},\mathbf{2},%
\mathbf{2}\right) }},
\end{array}
\notag \\
&&  \label{UCLA1}
\end{eqnarray}
yielding a non-BPS $Z=0$ mass splitting ``$14$ $m\neq 0$/$16$ $m=0$'' of the
vector multiplets' scalar degrees of freedom, matching the result obtained
in \cite{BFGM1}.\medskip

The model $stu$ is the one with the smallest number of vector multiplets ($%
n_{V}=3$) still exhibiting non-BPS $Z=0$ critical points. Without loss of
generality (due to \textit{triality symmetry}), one can identify $SU(2)_{%
\mathcal{R}}$ with the fourth factor $SU(2)$ in $H_{H}=SO(4)\otimes
SO(4)=\left( SU(2)\right) ^{4}$, whereas the $\mathcal{N}=2$ $\mathcal{R}$%
-symmetry can be identified with the third factor $SU(2)$ in $H_{H}$. Thus,
as yielded by Table 3, the $\mathcal{N}=2$ non-BPS $Z=0$ symmetry $%
\widetilde{h}^{\prime }\otimes H_{H}$ can be rewritten as
\begin{equation}
stu:\widetilde{h}^{\prime }\otimes H_{H}=SO(2)\otimes \left( SU(2)\right)
^{2}\otimes SU(2)_{H}\otimes SU(2)_{\mathcal{R}}.
\end{equation}
Thus, it holds that $SO(2)\otimes \left( SU(2)\right) ^{2}\otimes
SU(2)_{H}\subset SU(6)$.

Thus, by also recalling Eq. (\ref{BPS-mass-splitting}), one obtains that at $%
\mathcal{N}=2$ non-BPS $Z=0$ critical points the $\mathcal{N}=8$, $\frac{1}{8%
}$-BPS enhanced symmetry $SU(6)\otimes SU(2)_{\mathcal{R}}$ decomposes under
$\left( SU(2)\right) ^{2}\otimes SU(2)_{H}\otimes SU(2)_{\mathcal{R}}$ as
follows:
\begin{equation}
\begin{array}{l}
m\neq 0:\left( \mathbf{15},\mathbf{1}\right) \oplus \left( \overline{\mathbf{%
15}},\mathbf{1}\right) =6\left( \mathbf{1},\mathbf{1},\mathbf{1},\mathbf{1}%
\right) \oplus 2\left( \mathbf{2},\mathbf{2},\mathbf{1},\mathbf{1}\right)
\oplus 2\left( \mathbf{2},\mathbf{1},\mathbf{2},\mathbf{1}\right) \oplus
2\left( \mathbf{1},\mathbf{2},\mathbf{2},\mathbf{1}\right) ; \\
\\
m=0:\left( \mathbf{20},\mathbf{2}\right) =\left( \mathbf{2},\mathbf{2},%
\mathbf{2},\mathbf{2}\right) \oplus 2\left( \mathbf{1},\mathbf{1},\mathbf{2},%
\mathbf{2}\right) \oplus 2\left( \mathbf{1},\mathbf{2},\mathbf{1},\mathbf{2}%
\right) \oplus 2\left( \mathbf{2},\mathbf{1},\mathbf{1},\mathbf{2}\right) .
\end{array}
\label{UCLA2}
\end{equation}
Such a representation decomposition yields that at $\mathcal{N}=2$ non-BPS $%
Z=0$ critical points the vector multiplets' scalars and hypermultiplets'
scalars respectively sit in the following representations of $\left(
SU(2)\right) ^{2}\otimes SU(2)_{H}\otimes SU(2)_{\mathcal{R}}$:
\begin{eqnarray}
&&
\begin{array}{l}
\underset{\text{(}6\text{ in the }\mathcal{N}=2\text{, }d=4\text{ spectrum, }%
24\text{ truncated away) }}{30\text{ (real) vectors' scalar degrees of
freedom}}~=~\left\{
\begin{array}{l}
\underset{6\text{ in the }\mathcal{N}=2\text{, }d=4\text{ spectrum}}{%
\overset{m\neq 0\text{ }}{\overbrace{6\left( \mathbf{1},\mathbf{1},\mathbf{1}%
,\mathbf{1}\right) }}}\oplus \\
\\
\oplus \underset{24\text{ truncated away}}{\overset{m\neq 0}{\overbrace{%
2\left( \mathbf{2},\mathbf{2},\mathbf{1},\mathbf{1}\right) }}\oplus \overset{%
m=0}{\overbrace{2\left( \mathbf{1},\mathbf{2},\mathbf{1},\mathbf{2}\right)
\oplus 2\left( \mathbf{2},\mathbf{1},\mathbf{1},\mathbf{2}\right) }}};
\end{array}
\right. \\
\\
\underset{\text{(}16\text{ in the }\mathcal{N}=2\text{, }d=4\text{ spectrum,
}24\text{ truncated away)}}{40\text{ (real) hypers' scalar degrees of freedom%
}}~=~\left\{
\begin{array}{l}
\underset{16\text{ in the }\mathcal{N}=2\text{, }d=4\text{ spectrum}}{%
\overset{m=0}{\overbrace{\left( \mathbf{2},\mathbf{2},\mathbf{2},\mathbf{2}%
\right) }}}\oplus ~ \\
\\
\oplus \underset{24\text{ truncated away}}{\overset{m=0}{\overbrace{2\left(
\mathbf{1},\mathbf{1},\mathbf{2},\mathbf{2}\right) }}\oplus \overset{m\neq 0%
}{~\overbrace{2\left( \mathbf{2},\mathbf{1},\mathbf{2},\mathbf{1}\right)
\oplus 2\left( \mathbf{1},\mathbf{2},\mathbf{2},\mathbf{1}\right) }}},
\end{array}
\right.
\end{array}
\notag \\
&&
\end{eqnarray}
yielding that the Hessian of $V_{BH,\mathcal{N}=2}$ has no ``flat''
directions at non-BPS $Z=0$ critical points in the $stu$ model. As mentioned
above, this can be traced back to the noteworthy \textit{triality symmetry}
of the model under consideration, putting non-BPS $Z=0$ critical points on
the very same footing of $\frac{1}{2}$-BPS critical points.

Thus, in this sense one can state that in the $stu$ model the stability of $%
\frac{1}{2}$-BPS critical points implies, by \textit{triality symmetry}, the
stability of non-BPS $Z=0$ critical points. This can be quantitatively
understood by considering the representation decomposition of $SU(6)\otimes
SU(2)_{\mathcal{R}}$ in the $\frac{1}{2}$-BPS case. In such a case $SU(2)_{%
\mathcal{R}}=SU(2)_{H}$, and $SU(6)\otimes SU(2)_{\mathcal{R}}$ decomposes
into $H_{0}\otimes \frac{H_{H}}{SU(2)_{\mathcal{R}}}\otimes SU(2)_{\mathcal{R%
}}=\left( U(1)\right) ^{2}\otimes \left( SU(2)\right) ^{3}\otimes SU(2)_{%
\mathcal{R}}$ (once again, the choice of $SU(2)_{\mathcal{R}}$ as the fourth
$SU(2)$ does not imply any loss of generality, due to \textit{triality
symmetry}). It is thus easy to realize that this amounts simply to
interchange the third and fourth $SU(2)$s in the representation
decomposition (\ref{UCLA2}).

\section{\label{Sect5}$\mathcal{N}=8$ non-BPS Critical Points\newline
and\newline
$\mathcal{N}=2$ non-BPS $Z\neq 0$ Critical Points}

For the $\mathcal{N}=2$, $d=4$ supergravities listed in Table 1, the overall
symmetry $\mathcal{S}_{non-BPS,Z\neq 0}$ of $\mathcal{N}=2$, $d=4$ non-BPS $%
Z\neq 0$ critical points of $V_{BH,\mathcal{N}=2}$ is given by \cite{BFGM1}
\begin{equation}
\mathcal{S}_{non-BPS,Z\neq 0}=\widehat{h}\otimes H_{H},
\end{equation}
where $\widehat{h}$ is the \textit{m.c.s.} of the stabylizer $\widehat{H}$
of the $\mathcal{N}=2$ non-BPS $Z\neq 0$-supporting BH charge orbit \cite
{BFGM1}. Furthermore, $\mathcal{N}=2$ non-BPS $Z\neq 0$ case has $\mathcal{N}%
=2$ quartic $G_{V}$-invariant $I_{4}<0$. Thus, it is clear that $\mathcal{N}%
=2$ non-BPS $Z\neq 0$ case comes from the $\mathcal{N}=8\longrightarrow
\mathcal{N}=2$ supersymmetry reduction given by Eq. (\ref{SU(8)-->USp(8)}).
Thus, $\mathcal{S}_{non-BPS,Z\neq 0}$ must be included in the overall
enhanced symmetry $USp(8)$ of the $\mathcal{N}=8$ non-BPS case:
\begin{equation}
\mathcal{S}_{non-BPS,Z\neq 0}\subsetneq USp(8).  \label{UCLA3}
\end{equation}
It is worth pointing out that at $\mathcal{N}=2$ non-BPS $Z\neq 0$ critical
points of $V_{BH,\mathcal{N}=2}$ the group $SU(2)_{\mathcal{R}}$ cannot be
defined, and in general the $\mathcal{N}=2$ $\mathcal{R}$-symmetry $%
SU(2)_{H}\subsetneq H_{H}$, with the exception of the model $J_{3}^{\mathbb{H%
}}$, in which $n_{H}=0$ and thus $H_{H}$ cannot be defined and $%
SU(2)_{H}=G_{H}$ is a global symmetry.

In order to determine the mass degeneracy pattern of the Hessian of $V_{BH,%
\mathcal{N}=2}$ at $\mathcal{N}=2$ non-BPS $Z\neq 0$ critical points, one
will thus have to consider the decomposition of the representations $\mathbf{%
42}$ ($m=0$), $\mathbf{27}$ ($m\neq 0$) and $\mathbf{1}$ ($m\neq 0$) of the
enhanced $\mathcal{N}=8$ non-BPS symmetry $USp(8)$ (recall Eqs. (\ref
{SU(8)-->USp(8)}) and (\ref{non-BPS-Z<>0-mass-splitting})) into suitable
representations of $\mathcal{S}_{non-BPS,Z\neq 0}$. The embedding (\ref
{UCLA3}) is apriori not unique, but only one embedding among the possible
ones is consistent with the known quantum numbers of the vector and hyper
multiplets' scalars in the consider models, and thus consistent with the
performed supersymmetry reduction $\mathcal{N}=8\longrightarrow \mathcal{N}%
=2 $.

The corresponding data for all the $\mathcal{N}=2$, $d=4$ supergravities
which are consistent truncations of the $\mathcal{N}=8$, $d=4$ theory
(listed in Table 1) are given in Table 4 (for the column ``$\widehat{h}$''
refer to Table 8 of \cite{BFGM1}).
\begin{table}[t]
\begin{center}
\begin{tabular}{|c||c|c|c|}
\hline
& $
\begin{array}{c}
\\
\text{non-BPS, }Z\neq 0\text{ orbit} \\
\mathcal{O}_{non-BPS,Z\neq 0}=\frac{G_{V}}{\widehat{H}}~ \\
~
\end{array}
$ & $
\begin{array}{c}
\\
\widehat{h}\equiv m.c.s.\left( \widehat{H}\right) \\
~
\end{array}
$ & $
\begin{array}{c}
\\
\frac{H_{H}}{SU(2)_{H}} \\
~
\end{array}
$ \\ \hline\hline
$
\begin{array}{c}
\\
J_{3}^{\mathbb{H}} \\
~
\end{array}
$ & $\frac{SO^{\ast }(12)}{SU^{\ast }(6)}~$ & $
\begin{array}{c}
\\
USp(6) \\
~
\end{array}
$ & $\nexists H_{H},~~SU_{H}(2)=G_{H}~$ \\ \hline
$
\begin{array}{c}
\\
J_{3}^{\mathbb{C}} \\
~
\end{array}
$ & $\frac{SU(3,3)}{SL(3,\mathbb{C})}~$ & $
\begin{array}{c}
\\
SU(3) \\
~
\end{array}
~$ & $U(1)$ \\ \hline
$
\begin{array}{c}
\\
J_{3}^{\mathbb{R}} \\
~
\end{array}
$ & $\frac{Sp(6,\mathbb{R})}{SL(3,\mathbb{R})}$ & $
\begin{array}{c}
\\
SU(2) \\
~
\end{array}
$ & $SU(2)~$ \\ \hline
$
\begin{array}{c}
\\
stu \\
~
\end{array}
$ & $\frac{\left( SU(1,1)\right) ^{3}}{\left( SO(1,1)\right) ^{2}}~$ & $
\begin{array}{c}
\\
\mathbb{I} \\
~
\end{array}
~$ & $\left( SU(2)\right) ^{3}~$ \\ \hline
$
\begin{array}{c}
\\
J_{3,M}^{\mathbb{R}} \\
~
\end{array}
$ & $SU(1,1)$ & $
\begin{array}{c}
\\
\mathbb{I} \\
~
\end{array}
$ & $
\begin{array}{c}
\\
USp(6) \\
~
\end{array}
~$ \\ \hline
$
\begin{array}{c}
\\
J_{3,M}^{\mathbb{C}} \\
~
\end{array}
$ & $-$ & $-$ & $
\begin{array}{c}
\\
SU(6) \\
~
\end{array}
$ \\ \hline
\end{tabular}
\end{center}
\caption{\textbf{The non-BPS }$Z\neq 0$\textbf{\ supporting BH charge orbit }%
$\mathcal{O}_{non-BPS,Z\neq 0}$\textbf{, and the} \textbf{compact groups }$%
\widehat{h}$ \textbf{and }$\frac{H_{H}}{SU(2)_{H}}$ \textbf{(relevant at }$%
\mathcal{N}=2$\textbf{\ non-BPS }$Z\neq 0$ \textbf{critical points) for the}
$\mathcal{N}=2$\textbf{,} $d=4$ \textbf{supergravities listed in Table 1}}
\end{table}

In the following Subsects. we will analyze each model separately.

\subsection{\label{Subsect5-1}$J_{3}^{\mathbb{H}}$}

As given by Table 1, this model has $\left( n_{V},n_{H}\right) =\left(
15,0\right) $, and $\frac{G_{V}}{H_{V}}=\frac{SO^{\ast }(12)}{U(6)}$. $H_{H}$
cannot be defined, and $SU(2)_{H}=G_{H}$ is the global symmetry due to $%
n_{H}=0$. From Table 2 of \cite{ADF} the fundamental representation $\mathbf{%
56}$ of $G=E_{7(7)}$ decomposes along $G_{V}\otimes G_{H}=SO^{\ast
}(12)\otimes SU(2)_{H}$ as follows:
\begin{equation}
\mathbf{56}\longrightarrow \left( \mathbf{32},\mathbf{1}\right) \oplus
\left( \mathbf{12},\mathbf{2}\right) ,
\end{equation}
yielding that the $32$ real electric and magnetic charges $\left\{
p^{0},p^{1},....,p^{15},q_{0},q_{1},...q_{15}\right\} $ of the $1+15$
vectors of $J_{3}^{\mathbb{H}}$ lie in the $SU(2)_{H}$-singlet real
representation $\left( \mathbf{32},\mathbf{1}\right) $ of $SO^{\ast
}(12)\otimes SU(2)_{H}$ (here and in what follows the index ``$0$'' pertains
to the graviphoton). On the other hand, the fundamental representation $%
\mathbf{8}$ of the enhanced $\mathcal{N}=8$ non-BPS symmetry $USp(8)$
decomposes along $\mathcal{S}_{non-BPS,Z\neq 0}=\widehat{h}\otimes
SU(2)_{H}=USp(6)\otimes SU(2)_{H}$ as follows:
\begin{equation}
\mathbf{8}\longrightarrow \left( \mathbf{6},\mathbf{1}\right) \oplus \left(
\mathbf{1},\mathbf{2}\right) .
\end{equation}

The decomposition of the representations $\mathbf{42}$, $\mathbf{27}$ and $%
\mathbf{1}$ of $USp(8)$ along $\mathcal{S}_{non-BPS,Z\neq 0}$ and its
interpretation in terms of the $\mathcal{N}=2$, $d=4$ spectrum (and of the
truncated scalar degrees of freedom) reads as follows:
\begin{eqnarray}
&&
\begin{array}{l}
m=0:\mathbf{42}\longrightarrow \left\{
\begin{array}{l}
\overset{28\text{ }m=0\text{ hypers' scalar degrees of freedom truncated away%
}}{\overbrace{\left( \mathbf{14}^{\prime },\mathbf{2}\right) }}~\oplus \\
\\
\oplus \overset{14\text{ }m=0\text{ vectors' scalar degrees of freedom in }%
\mathcal{N}=2\text{, }d=4\text{ spectrum }}{~\overbrace{\left( \mathbf{14},%
\mathbf{1}\right) }};
\end{array}
\right. \\
\\
\\
\\
m\neq 0:\left\{
\begin{array}{l}
\mathbf{27}\longrightarrow \left\{
\begin{array}{l}
\overset{12\text{ }m\neq 0\text{ hypers' scalar degrees of freedom truncated
away}}{\overbrace{\left( \mathbf{6},\mathbf{2}\right) }}~\oplus ~ \\
\\
\oplus \overset{15\text{ }m\neq 0\text{ vectors' scalar degrees of freedom
in }\mathcal{N}=2\text{, }d=4\text{ spectrum}}{\overbrace{\left( \mathbf{1},%
\mathbf{1}\right) ~\oplus ~\left( \mathbf{14},\mathbf{1}\right) }};
\end{array}
\right. \\
\\
\\
\mathbf{1}\longrightarrow \overset{1\text{ }m\neq 0\text{ vectors' scalar
degree of freedom in }\mathcal{N}=2\text{, }d=4\text{ spectrum}}{\overbrace{%
\left( \mathbf{1},\mathbf{1}\right) }},
\end{array}
\right.
\end{array}
\notag  \label{UCLA-23-1} \\
&&
\end{eqnarray}
where $\mathbf{14}$ and $\mathbf{14}^{\prime }$ respectively stand for the
two-fold and three-fold antisymmetric (traceless) of $USp(6)$.

It should be noticed that for $J_{3}^{\mathbb{H}}$ the embedding of $%
\mathcal{S}_{non-BPS,Z\neq 0}$ in the enhanced $\mathcal{N}=8$ non-BPS
symmetry $USp(8)$ is unique. Moreover, since $J_{3}^{\mathbb{H}}$ has the
highest number $n_{V}=15$ of Abelian vector multiplets, all (would-be $%
\mathcal{N}=2$) vectors' scalar degrees of freedom of the starting $\mathcal{%
N}=8$ theory survive after the reduction $\mathcal{N}=8\longrightarrow
\mathcal{N}=2$.

The $\mathcal{N}=2$ non-BPS $Z\neq 0$ mass degeneracy pattern of the vector
multiplets' scalar degrees of freedom resulting from the decomposition (\ref
{UCLA-23-1}) is ``$n_{V}+1=16$ $m\neq 0$ / $n_{V}-1=14$ $m=0$'', thus
confirming the Hessian splitting found in \cite{TT}.

\subsection{\label{Subsect5-2}$J_{3}^{\mathbb{C}}$}

As given by Table 1, this model has $\left( n_{V},n_{H}\right) =\left(
9,1\right) $, and $\frac{G_{V}}{H_{V}}\otimes \frac{G_{H}}{H_{H}}=\frac{%
SU(3,3)}{SU(3)\otimes SU(3)\otimes U(1)}\otimes \frac{SU(2,1)}{%
SU(2)_{H}\otimes U(1)}$. From Table 2 of \cite{ADF} the fundamental
representation $\mathbf{56}$ of $G=E_{7(7)}$ decomposes along $G_{V}\otimes
G_{H}=SU(3,3)\otimes SU(2,1)$ as follows:
\begin{equation}
\mathbf{56}\longrightarrow \left( \mathbf{20},\mathbf{1}\right) \oplus
\left( \mathbf{6},\mathbf{3}\right) \oplus \left( \overline{\mathbf{6}},%
\overline{\mathbf{3}}\right) ,
\end{equation}
yielding that the $20$ real electric and magnetic charges $\left\{
p^{0},p^{1},....,p^{9},q_{0},q_{1},...q_{9}\right\} $ of the $1+9$ vectors
of $J_{3}^{\mathbb{C}}$ lie in the $SU(2,1)$-singlet real representation $%
\left( \mathbf{20},\mathbf{1}\right) $ of $SU(3,3)\otimes SU(2,1)$. On the
other hand, the fundamental representation $\mathbf{8}$ of the enhanced $%
\mathcal{N}=8$ non-BPS symmetry $USp(8)$ decomposes along $\mathcal{S}%
_{non-BPS,Z\neq 0}=\widehat{h}\otimes H_{H}=SU(3)\otimes SU(2)_{H}\otimes
U(1)$ as follows (here and in what follows we disregard the quantum numbers
of $U(1)$, not essential for our purposes):
\begin{equation}
\mathbf{8}\longrightarrow \left( \mathbf{3},\mathbf{1}\right) \oplus \left(
\overline{\mathbf{3}},\mathbf{1}\right) \oplus \left( \mathbf{1},\mathbf{2}%
\right) .
\end{equation}

The decomposition of the representations $\mathbf{42}$, $\mathbf{27}$ and $%
\mathbf{1}$ of $USp(8)$ along $\mathcal{S}_{non-BPS,Z\neq 0}$ and its
interpretation in terms of the $\mathcal{N}=2$, $d=4$ spectrum (and of the
truncated scalar degrees of freedom) reads as follows:
\begin{eqnarray}
&&
\begin{array}{l}
m=0:\mathbf{42}\longrightarrow \left\{
\begin{array}{l}
\overset{4\text{ }m=0\text{ hypers' scalar degrees of freedom in }\mathcal{N}%
=2\text{, }d=4\text{ spectrum}}{\overbrace{\left( \mathbf{1},\mathbf{2}%
\right) \oplus \left( \mathbf{1},\mathbf{2}\right) }}~\oplus \\
\\
\oplus ~\overset{6\text{ }m=0\text{ vectors' scalar degrees of freedom
truncated away }}{~\overbrace{\left( \overline{\mathbf{3}},\mathbf{1}\right)
\oplus \left( \mathbf{3},\mathbf{1}\right) }}~\oplus \\
\\
\oplus \overset{24\text{ }m=0\text{ hypers' scalar degrees of freedom
truncated away}}{\overbrace{\left( \overline{\mathbf{6}},\mathbf{2}\right)
\oplus \left( \mathbf{6},\mathbf{2}\right) }}~\oplus \\
\\
\oplus ~\overset{8\text{ }m=0\text{ vectors' scalar degrees of freedom in }%
\mathcal{N}=2\text{, }d=4\text{ spectrum }}{~\overbrace{\left( \mathbf{8},%
\mathbf{1}\right) }};
\end{array}
\right. \\
\\
\\
\\
m\neq 0:\left\{
\begin{array}{l}
\mathbf{27}\longrightarrow \left\{
\begin{array}{l}
\overset{6\text{ }m\neq 0\text{ vectors' scalar degrees of freedom truncated
away}}{\overbrace{\left( \overline{\mathbf{3}},\mathbf{1}\right) \oplus
\left( \mathbf{3},\mathbf{1}\right) }}~\oplus \\
\oplus ~\overset{8\text{ }m\neq 0\text{ vectors' scalar degrees of freedom
in }\mathcal{N}=2\text{, }d=4\text{ spectrum}}{\overbrace{\left( \mathbf{8},%
\mathbf{1}\right) }}~\oplus \\
\\
\overset{12\text{ }m\neq 0\text{ hypers' scalar degrees of freedom truncated
away}}{\overbrace{\left( \overline{\mathbf{3}},\mathbf{2}\right) \oplus
\left( \mathbf{3},\mathbf{2}\right) }}~\oplus \\
\\
\oplus ~\overset{1\text{ }m\neq 0\text{ vectors' scalar degree of freedom in
}\mathcal{N}=2\text{, }d=4\text{ spectrum}}{\overbrace{\left( \mathbf{1},%
\mathbf{1}\right) ~}};
\end{array}
\right. \\
\\
\\
\\
\mathbf{1}\longrightarrow \overset{1\text{ }m\neq 0\text{ vectors' scalar
degree of freedom in }\mathcal{N}=2\text{, }d=4\text{ spectrum}}{\overbrace{%
\left( \mathbf{1},\mathbf{1}\right) }}.
\end{array}
\right.
\end{array}
\notag \\
&&  \label{UCLA-23-2}
\end{eqnarray}

It should be noticed that for $J_{3}^{\mathbb{C}}$ the embedding of $%
\mathcal{S}_{non-BPS,Z\neq 0}$ in the enhanced $\mathcal{N}=8$ non-BPS
symmetry $USp(8)$ is apriori not unique, but the only consistent with the $%
\mathcal{N}=8\longrightarrow \mathcal{N}=2$ reduction originating $J_{3}^{%
\mathbb{C}}$ is the following two-step one:
\begin{equation}
USp(8)\supsetneq USp(6)\otimes USp(2)\supsetneq SU(3)\otimes
SU(2)_{H}\otimes U(1).
\end{equation}
Moreover, as evident from the decomposition (\ref{UCLA-23-2}), the $\mathcal{%
N}=8\longrightarrow \mathcal{N}=2$ reduction originating $J_{3}^{\mathbb{C}}$
truncates away:

1) $6$ $m=0$ and $6$ $m\neq 0$ vectors' scalar degrees of freedom, both sets
sitting in the $\left( \overline{\mathbf{3}},\mathbf{1}\right) \oplus \left(
\mathbf{3},\mathbf{1}\right) $ of $SU(3)\otimes SU(2)_{H}$;

2) $24$ $m=0$ and $12$ $m\neq 0$ hypers' scalar degrees of freedom,
respectively sitting in the $\left( \overline{\mathbf{6}},\mathbf{2}\right)
\oplus \left( \mathbf{6},\mathbf{2}\right) $ and $\left( \overline{\mathbf{3}%
},\mathbf{2}\right) \oplus \left( \mathbf{3},\mathbf{2}\right) $ of $%
SU(3)\otimes SU(2)_{H}$.

The resulting $\mathcal{N}=2$ $J_{3}^{\mathbb{C}}$ spectrum is composed by $%
4 $ $m=0$ real hypers' scalar degrees of freedom (rearranging in $1$
quaternionic hypermultiplet scalar), and by $n_{V}+1=10$ $m\neq 0$ and $%
n_{V}-1=8$ $m=0$ real vectors' scalar degrees of freedom, whose mass
degeneracy pattern thus confirms the Hessian splitting found in \cite{TT}.

\subsection{\label{Subsect5-3}$J_{3}^{\mathbb{R}}$}

As given by Table 1, this model has $\left( n_{V},n_{H}\right) =\left(
6,2\right) $, and $\frac{G_{V}}{H_{V}}\otimes \frac{G_{H}}{H_{H}}=\frac{Sp(6,%
\mathbb{R})}{U(3)}\otimes \frac{G_{2(2)}}{SU(2)\otimes SU(2)_{H}}$. From
Table 2 of \cite{ADF} the fundamental representation $\mathbf{56}$ of $%
G=E_{7(7)}$ decomposes along $G_{V}\otimes G_{H}=Sp(6,\mathbb{R})\otimes
G_{2(2)}$ as follows:
\begin{equation}
\mathbf{56}\longrightarrow \left( \mathbf{14}^{\prime },\mathbf{1}\right)
\oplus \left( \mathbf{6},\mathbf{7}\right) ,  \label{UCLA-23-3}
\end{equation}
where $\mathbf{14}^{\prime }$\ is the three-fold antisymmetric (traceless)
representation\ of $Sp(6,\mathbb{R})$. The decomposition (\ref{UCLA-23-3})
yields that the $14$ real electric and magnetic charges $\left\{
p^{0},p^{1},....,p^{6},q_{0},q_{1},...q_{6}\right\} $ of the $1+6$ vectors
of $J_{3}^{\mathbb{R}}$ lie in the $G_{2(2)}$-singlet real representation $%
\left( \mathbf{14}^{\prime },\mathbf{1}\right) $ of $Sp(6,\mathbb{R})\otimes
G_{2(2)}$. The symmetry group $\mathcal{S}_{non-BPS,Z\neq 0}$ of $J_{3}^{%
\mathbb{R}}$ reads
\begin{equation}
\mathcal{S}_{non-BPS,Z\neq 0}=\widehat{h}\otimes H_{H}=SO(3)\otimes
SU(2)\otimes SU(2).
\end{equation}
Thus, apriori $\mathcal{S}_{non-BPS,Z\neq 0}$ can be embedded in the
enhanced $\mathcal{N}=8$ non-BPS symmetry $USp(8)$ in many ways, but the
only consistent with the $\mathcal{N}=8\longrightarrow \mathcal{N}=2$
reduction originating $J_{3}^{\mathbb{R}}$ is the following two-step one:
\begin{equation}
USp(8)\supsetneq \underset{
\begin{array}{c}
\cup \\
SU(2)_{P}
\end{array}
}{USp(4)}\underset{
\begin{array}{c}
~ \\
\otimes
\end{array}
}{\otimes }\underset{
\begin{array}{c}
\cup \\
SU(2)\otimes SU(2)_{H},
\end{array}
}{USp(4)},~~H_{H}=SU(2)_{P}\otimes SU(2)_{H},  \label{UCLA-23-4}
\end{equation}
yielding that $\mathcal{S}_{non-BPS,Z\neq 0}$ can be rewritten as
\begin{equation}
\mathcal{S}_{non-BPS,Z\neq 0}=SU(2)_{P}\otimes SU(2)\otimes SU(2)_{H},
\end{equation}
where $SU(2)_{P}=\frac{H_{H}}{SU(2)_{H}}$ is the $SU(2)$-\textit{principal
embedding}\footnote{%
The group sequence $USp(n)_{n\in \mathbb{N}}$ has an embedding, called
\textit{principal}, in $SU(2)$ with spin $s_{n}=n-\frac{1}{2}$ \cite{Slansky}%
.} of one (say, without any loss of generality, of the first) of the two $%
USp(4)$, thus sitting in a spin $s=\frac{3}{2}$ representation $\left(
\mathbf{4},\mathbf{1},\mathbf{1}\right) $ with respect to $SU(2)_{P}\otimes
SU(2)\otimes SU(2)_{H}$. The identification $H_{H}=SU(2)_{P}\otimes
SU(2)_{H} $ is consistent with the known result that the hypermultiplets'
quaternionic scalars of $J_{3}^{\mathbb{R}}$ have spins $\left( s,s^{\prime
}\right) =\left( \frac{3}{2},\frac{1}{2}\right) $ with respect to $H_{H}$,
and thus sit in a representation $\left( \mathbf{4},\mathbf{2}\right) $ of
such a stabylizer, where the spin $s^{\prime }=\frac{1}{2}$ is with respect
to the $\mathcal{N}=2$ $\mathcal{R}$-symmetry $SU(2)_{H}$ in $H_{H}$. Thus,
the fundamental representation $\mathbf{8}$ of the enhanced $\mathcal{N}=8$
non-BPS symmetry $USp(8)$ decomposes along $SU(2)_{P}\otimes SU(2)\otimes
SU(2)_{H}$ as follows:
\begin{equation}
\mathbf{8}\longrightarrow \left( \mathbf{1},\mathbf{2},\mathbf{1}\right)
\oplus \left( \mathbf{4},\mathbf{1},\mathbf{1}\right) \oplus \left( \mathbf{1%
},\mathbf{1},\mathbf{2}\right) .
\end{equation}

The decomposition of the representations $\mathbf{42}$, $\mathbf{27}$ and $%
\mathbf{1}$ of $USp(8)$ along $\mathcal{S}_{non-BPS,Z\neq 0}$ and its
interpretation in terms of the $\mathcal{N}=2$, $d=4$ spectrum (and of the
truncated scalar degrees of freedom) reads as follows:
\begin{eqnarray}
&&
\begin{array}{l}
m=0:\mathbf{42}\longrightarrow \left\{
\begin{array}{l}
\overset{5\text{ }m=0\text{ vectors' scalar degrees of freedom in }\mathcal{N%
}=2\text{, }d=4\text{ spectrum}}{\overbrace{\left( \mathbf{5},\mathbf{1},%
\mathbf{1}\right) }}~\oplus \\
\\
\oplus \overset{20\text{ }m=0\text{ hypers' scalar degrees of freedom
truncated away }}{~\overbrace{\left( \mathbf{5},\mathbf{2},\mathbf{2}\right)
}}~\oplus \\
\\
\oplus \overset{8\text{ }m=0\text{ hypers' scalar degrees of freedom in }%
\mathcal{N}=2\text{, }d=4\text{ spectrum}}{\overbrace{\left( \mathbf{4},%
\mathbf{1},\mathbf{2}\right) }}~\oplus \\
\\
\oplus \overset{9\text{ }m=0\text{ vectors' scalar degrees of freedom
truncated away }}{~\overbrace{\left( \mathbf{4},\mathbf{2},\mathbf{1}\right)
\oplus \left( \mathbf{1},\mathbf{1},\mathbf{1}\right) }};
\end{array}
\right. \\
\\
\\
\\
m\neq 0:\left\{
\begin{array}{l}
\mathbf{27}\longrightarrow \left\{
\begin{array}{l}
\overset{6\text{ }m\neq 0\text{ vectors' scalar degrees of freedom in }%
\mathcal{N}=2\text{, }d=4\text{ spectrum}}{\overbrace{\left( \mathbf{5},%
\mathbf{1},\mathbf{1}\right) \oplus \left( \mathbf{1},\mathbf{1},\mathbf{1}%
\right) }}~\oplus \\
\\
\oplus ~\overset{9\text{ }m=0\text{ vectors' scalar degrees of freedom
truncated away }}{\overbrace{\left( \mathbf{4},\mathbf{2},\mathbf{1}\right)
\oplus \left( \mathbf{1},\mathbf{1},\mathbf{1}\right) }}~\oplus \\
\\
\overset{12\text{ }m\neq 0\text{ hypers' scalar degrees of freedom truncated
away}}{\overbrace{\left( \mathbf{4},\mathbf{1},\mathbf{2}\right) \oplus
\left( \mathbf{1},\mathbf{2},\mathbf{2}\right) }}~;
\end{array}
\right. \\
\\
\\
\\
\mathbf{1}\longrightarrow \overset{1\text{ }m\neq 0\text{ vectors' scalar
degree of freedom in }\mathcal{N}=2\text{, }d=4\text{ spectrum}}{\overbrace{%
\left( \mathbf{1},\mathbf{1},\mathbf{1}\right) }}.
\end{array}
\right.
\end{array}
\notag \\
&&
\end{eqnarray}

Such decompositions yield that the $\mathcal{N}=8\longrightarrow \mathcal{N}%
=2$ reduction originating $J_{3}^{\mathbb{R}}$ truncates away:

1) $9$ $m=0$ and $9$ $m\neq 0$ vectors' scalar degrees of freedom, both sets
sitting in the $\left( \mathbf{4},\mathbf{2},\mathbf{1}\right) \oplus \left(
\mathbf{1},\mathbf{1},\mathbf{1}\right) $ of $SU(2)_{P}\otimes SU(2)\otimes
SU(2)_{H}$;

2) $20$ $m=0$ and $12$ $m\neq 0$ hypers' scalar degrees of freedom,
respectively sitting in the $\left( \mathbf{5},\mathbf{2},\mathbf{2}\right) $
and $\left( \mathbf{4},\mathbf{1},\mathbf{2}\right) \oplus \left( \mathbf{1},%
\mathbf{2},\mathbf{2}\right) $ of $SU(2)_{P}\otimes SU(2)\otimes SU(2)_{H}$.

The resulting $\mathcal{N}=2$ $J_{3}^{\mathbb{R}}$ spectrum is composed by $%
8 $ $m=0$ real hypers' scalar degrees of freedom (rearranging in $2$
quaternionic hypermultiplet scalar), and by $n_{V}+1=7$ $m\neq 0$ and $%
n_{V}-1=5~m=0$ real vectors' scalar degrees of freedom, whose mass
degeneracy pattern thus confirms once again the Hessian splitting found in
\cite{TT}.

\subsection{\label{Subsect5-4}$stu$}

As given by Table 1, this model has $\left( n_{V},n_{H}\right) =\left(
3,4\right) $, and $\frac{G_{V}}{H_{V}}\otimes \frac{G_{H}}{H_{H}}=\frac{%
SU(1,1)}{U(1)}\otimes \frac{SO(2,2)}{SO(2)\otimes SO(2)}\otimes \frac{SO(4,4)%
}{SO(4)\otimes SO(4)}$. From Eq. (182) of \cite{ADFFT} the fundamental
representation $\mathbf{56}$ of $G=E_{7(7)}$ decomposes along $G_{V}\otimes
G_{H}=SU(1,1)\otimes SO(2,2)\otimes SO(4,4)\sim $ $\left( SU(1,1)\right)
^{3}\otimes SO(4,4)$ as follows (the three $SU(1,1)$\ are actually
indistinguishable due to \textit{triality symmetry}):
\begin{equation}
\mathbf{56}\longrightarrow \left( \mathbf{2},\mathbf{2},\mathbf{2},\mathbf{1}%
\right) \oplus \left( \mathbf{2},\mathbf{1},\mathbf{1},\mathbf{8}_{v}\right)
\oplus \left( \mathbf{1},\mathbf{2},\mathbf{1},\mathbf{8}_{s}\right) \oplus
\left( \mathbf{1},\mathbf{1},\mathbf{2},\mathbf{8}_{c}\right) ,
\label{UCLA-23-3-bis}
\end{equation}
where $\mathbf{8}_{v}$, $\mathbf{8}_{s}$\ and $\mathbf{8}_{c}$\ respectively
are the vector, chiral spinorial and anti-chiral spinorial representations
of $SO(4,4)$. The decomposition (\ref{UCLA-23-3-bis}) yields that the $8$
real electric and magnetic charges $\left\{
p^{0},p^{1},....,p^{3},q_{0},q_{1},...q_{3}\right\} $ of the $1+3$ vectors
of the $stu$ model lie in the $SO(4,4)$-singlet real representation $\left(
\mathbf{2},\mathbf{2},\mathbf{2},\mathbf{1}\right) $ of $\left(
SU(1,1)\right) ^{3}\otimes SO(4,4)$. The symmetry group $\mathcal{S}%
_{non-BPS,Z\neq 0}$ of the $stu$ model reads
\begin{equation}
\mathcal{S}_{non-BPS,Z\neq 0}=\widehat{h}\otimes H_{H}\overset{\widehat{h}%
_{stu}=\mathbb{I}}{=}H_{H}=SO(4)\otimes SO(4)\sim \left( SU(2)\right) ^{4}.
\end{equation}
Thus, apriori $\mathcal{S}_{non-BPS,Z\neq 0}$ can be embedded in the
enhanced $\mathcal{N}=8$ non-BPS symmetry $USp(8)$ in many ways, but the
only consistent with the $\mathcal{N}=8\longrightarrow \mathcal{N}=2$
reduction originating the $stu$ model is the following two-step one:
\begin{equation}
USp(8)\supsetneq USp(4)\otimes USp(4)\supsetneq SO(4)\otimes SO(4)\sim
\left( SU(2)\right) ^{4}.  \label{UCLA-23-night-2}
\end{equation}
We can choose the $\mathcal{N}=2$ $\mathcal{R}$-symmetry $SU(2)_{H}$ to be
the fourth one in $\mathcal{S}_{non-BPS,Z\neq 0}$ (as we will see below,
such an arbitrariness in the choice of the placement of the $\mathcal{N}=2$ $%
\mathcal{R}$-symmetry inside $H_{H}$ is actually removed by the \textit{%
triality symmetry} of the $stu$ model). Consequently, $\mathcal{S}%
_{non-BPS,Z\neq 0}$ can be rewritten as
\begin{equation}
\mathcal{S}_{non-BPS,Z\neq 0}=\left( SU(2)\right) ^{3}\otimes SU(2)_{H}.
\end{equation}
Thus, the fundamental representation $\mathbf{8}$ of the enhanced $\mathcal{N%
}=8$ non-BPS symmetry $USp(8)$ decomposes along the chain of branchings (\ref
{UCLA-23-night-2}) as follows:
\begin{equation}
\underset{USp(8)}{\mathbf{8}}\longrightarrow \underset{USp(4)\otimes USp(4)}{%
\left( \mathbf{4},\mathbf{1}\right) \oplus \left( \mathbf{1},\mathbf{4}%
\right) }\longrightarrow \underset{SO(4)\otimes SO(4)}{\left( \mathbf{4}_{s},%
\mathbf{1}\right) \oplus \left( \mathbf{1},\mathbf{4}_{s}\right) }%
\longrightarrow \underset{SU(2)\otimes SU(2)\otimes SU(2)\otimes SU(2)_{H}}{%
\left( \mathbf{2},\mathbf{1},\mathbf{1},\mathbf{1}\right) \oplus \left(
\mathbf{1},\mathbf{2},\mathbf{1},\mathbf{1}\right) \oplus \left( \mathbf{1},%
\mathbf{1},\mathbf{2},\mathbf{1}\right) \oplus \left( \mathbf{1},\mathbf{1},%
\mathbf{1},\mathbf{2}\right) },
\end{equation}
where $\mathbf{4}_{s}$ is the spinorial of $SO(4)$ (or, equivalently, the
reduction of the fundamental of $USp(4)$ with respect to $SO(4)$).

Due to the chain of group inclusions (\ref{UCLA-23-night-2}) needed in the $%
stu$ model in order to correctly embed $\mathcal{S}_{non-BPS,Z\neq 0}$ into $%
USp(8)$, the decomposition of the representations $\mathbf{42}$, $\mathbf{27}
$ and $\mathbf{1}$ of $USp(8)$ along $\mathcal{S}_{non-BPS,Z\neq 0}$ should
better be performed in two steps:

\textit{i}) decomposition of $USp(8)$ along $USp(4)\otimes USp(4)$. It
respectively yields (the prime distinguishes the - representations of the -
two $USp(4)$)
\begin{eqnarray}
&&
\begin{array}{l}
m=0:\mathbf{42\longrightarrow }\left( \mathbf{4},\mathbf{4}^{^{\prime
}}\right) \oplus \left( \mathbf{5},\mathbf{5}^{^{\prime }}\right) \oplus
\left( \mathbf{1},\mathbf{1}^{^{\prime }}\right) ; \\
\\
\\
m\neq 0:\left\{
\begin{array}{l}
\mathbf{27\longrightarrow }\left( \mathbf{4},\mathbf{4}^{^{\prime }}\right)
\oplus \left( \mathbf{5},\mathbf{1}^{^{\prime }}\right) \oplus \left(
\mathbf{1},\mathbf{5}^{^{\prime }}\right) \oplus \left( \mathbf{1},\mathbf{1}%
^{^{\prime }}\right) ; \\
\\
\mathbf{1\longrightarrow }\left( \mathbf{1},\mathbf{1}^{^{\prime }}\right) .
\end{array}
\right.
\end{array}
\notag \\
&&
\end{eqnarray}

\textit{ii}) Decomposition of $USp(4)\otimes USp(4)$ along $SO(4)\otimes
SO(4)$. It will involve the representations $\mathbf{4}_{s}$ (previously
introduced) and $\mathbf{4}_{v}$ (vector representation of $SO(4)$\ or,
equivalently, reduction of the antisymmetric traceless of $USp(4)$\textbf{\ }%
with respect to $SO(4)$). By exploiting the following decompositions of the
representations $\left( \mathbf{4},\mathbf{4}^{^{\prime }}\right) $, $\left(
\mathbf{5},\mathbf{5}^{^{\prime }}\right) $, $\left( \mathbf{5},\mathbf{1}%
^{^{\prime }}\right) $ and $\left( \mathbf{1},\mathbf{1}^{^{\prime }}\right)
$ of $USp(4)\otimes USp(4)$ along $SO(4)\otimes SO(4)$:
\begin{equation}
\begin{array}{l}
\left( \mathbf{4},\mathbf{4}^{^{\prime }}\right) \longrightarrow \left(
\mathbf{4}_{s},\mathbf{4}_{s}^{^{\prime }}\right) ; \\
\\
\left( \mathbf{5},\mathbf{5}^{^{\prime }}\right) \longrightarrow \left(
\mathbf{4}_{v},\mathbf{4}_{v}^{^{\prime }}\right) \oplus \left( \mathbf{4}%
_{v},\mathbf{1}^{^{\prime }}\right) \oplus \left( \mathbf{1},\mathbf{4}%
_{v}^{^{\prime }}\right) \oplus \left( \mathbf{1},\mathbf{1}^{^{\prime
}}\right) ; \\
\\
\left( \mathbf{5},\mathbf{1}^{^{\prime }}\right) \longrightarrow \left(
\mathbf{4}_{v},\mathbf{1}^{^{\prime }}\right) \oplus \left( \mathbf{1},%
\mathbf{1}^{^{\prime }}\right) ; \\
\\
\left( \mathbf{1},\mathbf{1}^{^{\prime }}\right) \longrightarrow \left(
\mathbf{1},\mathbf{1}^{^{\prime }}\right) ,
\end{array}
\end{equation}
one gets the following decompositions of representations $\mathbf{42}$, $%
\mathbf{27}$ and $\mathbf{1}$ of $USp(8)$ along $SO(4)\otimes SO(4)$:
\begin{eqnarray}
&&
\begin{array}{l}
m=0:\mathbf{42\longrightarrow }\left( \mathbf{4}_{s},\mathbf{4}%
_{s}^{^{\prime }}\right) \oplus \left( \mathbf{4}_{v},\mathbf{4}%
_{v}^{^{\prime }}\right) \oplus \left( \mathbf{4}_{v},\mathbf{1}^{^{\prime
}}\right) \oplus \left( \mathbf{1},\mathbf{4}_{v}^{^{\prime }}\right) \oplus
2\left( \mathbf{1},\mathbf{1}^{^{\prime }}\right) ; \\
\\
\\
m\neq 0:\left\{
\begin{array}{l}
\mathbf{27\longrightarrow }\left( \mathbf{4}_{s},\mathbf{4}_{s}^{^{\prime
}}\right) \oplus \left( \mathbf{4}_{v},\mathbf{1}^{^{\prime }}\right) \oplus
\left( \mathbf{1},\mathbf{4}_{v}^{^{\prime }}\right) \oplus 3\left( \mathbf{1%
},\mathbf{1}^{^{\prime }}\right) ; \\
\\
\mathbf{1\longrightarrow }\left( \mathbf{1},\mathbf{1}^{^{\prime }}\right) .
\end{array}
\right.
\end{array}
\notag \\
&&
\end{eqnarray}

\textit{iii}) Further decomposition, performed by exploiting the group
isomorphism $SO(4)\sim SU(2)\otimes SU(2)$. Under the group isomorphism $%
SO(4)\sim \left( SU(2)\right) ^{2}$, $\mathbf{4}_{s}$ and $\mathbf{4}_{v}$
respectively decompose as follows:
\begin{equation}
\begin{array}{l}
\mathbf{4}_{s}\longrightarrow \left( \mathbf{2},\mathbf{1}\right) \oplus
\left( \mathbf{1},\mathbf{2}\right) ; \\
\\
\mathbf{4}_{v}\longrightarrow \left( \mathbf{2},\mathbf{2}\right) .
\end{array}
\end{equation}
Thus, the decomposition of representations $\mathbf{42}$, $\mathbf{27}$ and $%
\mathbf{1}$ of $USp(8)$ along $\left( SU(2)\right) ^{4}=\left( SU(2)\right)
^{3}\otimes SU(2)_{H}$ (embedded into $USp(8)$ in the way given by the chain
(\ref{UCLA-23-night-2}) of group inclusions), and its interpretation in
terms of the $\mathcal{N}=2$, $d=4$ spectrum (and of the truncated scalar
degrees of freedom), reads as follows:
\begin{eqnarray}
&&
\begin{array}{l}
m=0:\mathbf{42\longrightarrow }\left\{
\begin{array}{l}
\overset{12\text{ }m=0\text{ vectors' scalar degrees of freedom truncated
away}}{\overbrace{\left( \mathbf{2},\mathbf{1},\mathbf{2}^{\prime },\mathbf{1%
}^{\prime }\right) \oplus \left( \mathbf{1},\mathbf{2},\mathbf{2}^{\prime },%
\mathbf{1}^{\prime }\right) \oplus ~\left( \mathbf{2},\mathbf{2},\mathbf{1}%
^{\prime },\mathbf{1}^{\prime }\right) }}\oplus \\
\\
\oplus ~\overset{12\text{ }m=0\text{ hypers' scalar degrees of freedom
truncated away}}{\overbrace{\left( \mathbf{1},\mathbf{2},\mathbf{1}^{\prime
},\mathbf{2}^{\prime }\right) \oplus \left( \mathbf{2},\mathbf{1},\mathbf{1}%
^{\prime },\mathbf{2}^{\prime }\right) \oplus \left( \mathbf{1},\mathbf{1},%
\mathbf{2}^{\prime },\mathbf{2}^{\prime }\right) }}~\oplus \\
\\
\oplus ~\overset{16\text{ }m=0\text{ hypers' scalar degrees of freedom in }%
\mathcal{N}=2\text{, }d=4\text{ spectrum}}{\overbrace{\left( \mathbf{2},%
\mathbf{2},\mathbf{2}^{\prime },\mathbf{2}^{\prime }\right) }~}~\oplus \\
\\
\oplus ~\overset{2\text{ }m=0\text{ vectors' scalar degrees of freedom in }%
\mathcal{N}=2\text{, }d=4\text{ spectrum}}{\overbrace{2\left( \mathbf{1},%
\mathbf{1},\mathbf{1}^{^{\prime }},\mathbf{1}^{^{\prime }}\right) }};
\end{array}
\right. \\
\\
\\
\\
m\neq 0:\left\{
\begin{array}{l}
\mathbf{27\longrightarrow }\left\{
\begin{array}{l}
\overset{12\text{ }m\neq 0\text{ vectors' scalar degrees of freedom
truncated away}}{\overbrace{\left( \mathbf{2},\mathbf{1},\mathbf{2}^{\prime
},\mathbf{1}^{\prime }\right) \oplus \left( \mathbf{1},\mathbf{2},\mathbf{2}%
^{\prime },\mathbf{1}^{\prime }\right) ~\oplus \left( \mathbf{2},\mathbf{2},%
\mathbf{1}^{\prime },\mathbf{1}^{\prime }\right) }}~\oplus \\
\\
\oplus ~\overset{12\text{ }m\neq 0\text{ hypers' scalar degrees of freedom
truncated away}}{\overbrace{\left( \mathbf{1},\mathbf{2},\mathbf{1}^{\prime
},\mathbf{2}^{\prime }\right) \oplus \left( \mathbf{2},\mathbf{1},\mathbf{1}%
^{\prime },\mathbf{2}^{\prime }\right) \oplus \left( \mathbf{1},\mathbf{1},%
\mathbf{2}^{\prime },\mathbf{2}^{\prime }\right) }~}\oplus \\
\\
\oplus ~\overset{3\text{ }m\neq 0\text{ vectors' scalar degrees of freedom
in }\mathcal{N}=2\text{, }d=4\text{ spectrum}}{\overbrace{3\left( \mathbf{1},%
\mathbf{1},\mathbf{1}^{^{\prime }},\mathbf{1}^{^{\prime }}\right) }};
\end{array}
\right. \\
\\
\\
\mathbf{1\longrightarrow }\overset{1\text{ }m\neq 0\text{ vectors' scalar
degree of freedom in }\mathcal{N}=2\text{, }d=4\text{ spectrum}}{\overbrace{%
\left( \mathbf{1},\mathbf{1},\mathbf{1}^{^{\prime }},\mathbf{1}^{^{\prime
}}\right) }}.
\end{array}
\right.
\end{array}
\notag \\
&&  \label{UCLA-24-1}
\end{eqnarray}
\newline

Such decompositions yield that the $\mathcal{N}=8\longrightarrow \mathcal{N}%
=2$ reduction originating the $stu$ model truncates away:

1) $12$ $m=0$ and $12$ $m\neq 0$ vectors' scalar degrees of freedom, both
sets sitting in the $\left( \mathbf{2},\mathbf{1},\mathbf{2}^{\prime },%
\mathbf{1}^{\prime }\right) \oplus \left( \mathbf{1},\mathbf{2},\mathbf{2}%
^{\prime },\mathbf{1}^{\prime }\right) ~\oplus \left( \mathbf{2},\mathbf{2},%
\mathbf{1}^{\prime },\mathbf{1}^{\prime }\right) $ of $\left( SU(2)\right)
^{3}\otimes SU(2)_{H}$ (note the \textit{triality symmetry} acting on the
first three quantum numbers);

2) $12$ $m=0$ and $12$ $m\neq 0$ hypers' scalar degrees of freedom, both
sets sitting in the $\left( \mathbf{1},\mathbf{2},\mathbf{1}^{\prime },%
\mathbf{2}^{\prime }\right) \oplus \left( \mathbf{2},\mathbf{1},\mathbf{1}%
^{\prime },\mathbf{2}^{\prime }\right) \oplus \left( \mathbf{1},\mathbf{1},%
\mathbf{2}^{\prime },\mathbf{2}^{\prime }\right) $ of $\left( SU(2)\right)
^{3}\otimes SU(2)_{H}$ (note the \textit{triality symmetry} acting on the
first three quantum numbers).

As it is seen,both the vectors' and hypers' scalar degrees of freedom
truncated out receive half of the contribution from the $\mathbf{42}$
(massless) of $USp(8)$ and the other half of the contribution from the $%
\mathbf{27}$ (massive) of $USp(8)$. As it holds in general, the massive
singlet representation $\mathbf{1}$ of $USp(8)$ always appears in the $%
\mathcal{N}=2$, $d=4$ resulting spectrum.

The spectrum of the $\mathcal{N}=2$, $d=4$ $stu$ model determined by the
decompositions (\ref{UCLA-24-1}) is composed by $16$ $m=0$ real hypers'
scalar degrees of freedom (rearranging in $4$ quaternionic hypermultiplet
scalar), and by $n_{V}+1=4$ $m\neq 0$ and $n_{V}-1=2~m=0$ real vectors'
scalar degrees of freedom, whose mass degeneracy pattern thus confirms once
again the Hessian splitting found in \cite{TT}.

\subsection{\label{Subsect5-5}$J_{3,M}^{\mathbb{R}}$}

As given by Table 1, this model has $\left( n_{V},n_{H}\right) =\left(
1,7\right) $, and $\frac{G_{V}}{H_{V}}\otimes \frac{G_{H}}{H_{H}}=\frac{%
SU(1,1)}{U(1)}\otimes \frac{F_{4(4)}}{USp(6)\otimes SU(2)_{H}}$ (recall that
$USp(2)\sim SU(2)$). From Table 2 of \cite{ADF} the fundamental
representation $\mathbf{56}$ of $G=E_{7(7)}$ decomposes along $G_{V}\otimes
G_{H}=SU(1,1)\otimes F_{4(4)}$ as follows:
\begin{equation}
\mathbf{56}\longrightarrow \left( \mathbf{4},\mathbf{1}\right) \oplus \left(
\mathbf{2},\mathbf{26}\right) .
\end{equation}
Such a decomposition yields that the $4$ real electric and magnetic charges $%
\left\{ p^{0},p^{1},q_{0},q_{1}\right\} $ of the $1+1$ vectors of $J_{3,M}^{%
\mathbb{R}}$ lie in the $F_{4(4)}$-singlet real representation $\left(
\mathbf{4},\mathbf{1}\right) $ of $SU(1,1)\otimes F_{4(4)}$. The
representation $\mathbf{4}$\ of $SU(1,1)$\ corresponds to spin $s=\frac{3}{2}
$, and this identifies\textbf{\ }$\frac{G_{V}}{H_{V}}=\frac{SU(1,1)}{U(1)}$\
as a special K\"{a}hler manifold ($dim_{\mathbb{C}}=1$) with cubic
holomorphic prepotential\textbf{\ }reading\footnote{%
For a discussion of (the $\mathcal{N}=2$, $d=4$ attractor Eqs. in the
special K\"{a}hler geometry of) $\frac{SU(1,1)}{U(1)}$ with cubic
holomorphic prepotential, see \textit{e.g.} \cite{BFGM1,BFM-SIGRAV06} (and
Refs. therein) and \cite{Saraikin-Vafa-1}.} (in a suitable system of special
projective coordinates) $\mathcal{F}\left( t\right) =\lambda t^{3}$, $%
\lambda \in \mathbb{C}_{0}$\textbf{.} The symmetry group $\mathcal{S}%
_{non-BPS,Z\neq 0}$ of $J_{3,M}^{\mathbb{R}}$ is the same of the one of $%
J_{3}^{\mathbb{H}}$, and it reads ($\widehat{h}=\mathbb{I}$, as in the $stu$
model)
\begin{equation}
\mathcal{S}_{non-BPS,Z\neq 0}=\widehat{h}\otimes H_{H}=H_{H}=USp(6)\otimes
SU(2)_{H}.
\end{equation}
As it holds also for $J_{3}^{\mathbb{H}}$, in the model $J_{3,M}^{\mathbb{R}%
} $ the embedding of $\mathcal{S}_{non-BPS,Z\neq 0}$ in the enhanced $%
\mathcal{N}=8$ non-BPS symmetry $USp(8)$ is unique. The fundamental
representation $\mathbf{8}$ of $USp(8)$ decomposes along $USp(6)\otimes
SU(2)_{H}$ as follows:
\begin{equation}
\mathbf{8}\longrightarrow \left( \mathbf{6},\mathbf{1}\right) \oplus \left(
\mathbf{1},\mathbf{2}\right) .
\end{equation}

The decomposition of the representations $\mathbf{42}$, $\mathbf{27}$ and $%
\mathbf{1}$ of $USp(8)$ along $\mathcal{S}_{non-BPS,Z\neq 0}$ and its
interpretation in terms of the $\mathcal{N}=2$, $d=4$ spectrum (and of the
truncated scalar degrees of freedom) reads as follows:
\begin{eqnarray}
&&
\begin{array}{l}
m=0:\mathbf{42}\longrightarrow \left\{
\begin{array}{l}
\overset{14\text{ }m=0\text{ vectors' scalar degrees of freedom truncated
away}}{\overbrace{\left( \mathbf{14},\mathbf{1}\right) }}~\oplus \\
\\
\oplus \overset{28\text{ }m=0\text{ hypers' scalar degrees of freedom in }%
\mathcal{N}=2\text{, }d=4\text{ spectrum}}{~\overbrace{\left( \mathbf{14}%
^{\prime },\mathbf{2}\right) }}~;
\end{array}
\right. \\
\\
\\
\\
m\neq 0:\left\{
\begin{array}{l}
\mathbf{27}\longrightarrow \left\{
\begin{array}{l}
\overset{12\text{ }m\neq 0\text{ hypers' scalar degrees of freedom truncated
away }}{\overbrace{\left( \mathbf{6},\mathbf{2}\right) }}~\oplus \\
\\
\oplus ~\overset{14\text{ }m\neq 0\text{ vectors' scalar degrees of freedom
truncated away }}{\overbrace{\left( \mathbf{14},\mathbf{1}\right) }}~\oplus
\\
\\
\overset{1\text{ }m\neq 0\text{ vectors' scalar degree of freedom in }%
\mathcal{N}=2\text{, }d=4\text{ spectrum}}{\overbrace{\left( \mathbf{1},%
\mathbf{1}\right) }}~;
\end{array}
\right. \\
\\
\\
\\
\mathbf{1}\longrightarrow \overset{1\text{ }m\neq 0\text{ vectors' scalar
degree of freedom in }\mathcal{N}=2\text{, }d=4\text{ spectrum}}{\overbrace{%
\left( \mathbf{1},\mathbf{1}\right) }},
\end{array}
\right.
\end{array}
\notag \\
&&
\end{eqnarray}
where $\mathbf{14}$ and $\mathbf{14}^{\prime }$ respectively stand for the
two-fold and three-fold antisymmetric (traceless) of $USp(6)$.

Such decompositions yield that the $\mathcal{N}=8\longrightarrow \mathcal{N}%
=2$ reduction originating $J_{3,M}^{\mathbb{R}}$ truncates away:

1) $14$ $m=0$ and $14$ $m\neq 0$ vectors' scalar degrees of freedom, both
sets sitting in the $\left( \mathbf{14},\mathbf{1}\right) $ of $%
USp(6)\otimes SU(2)_{H}$;

2) $12$ $m\neq 0$ hypers' scalar degrees of freedom, sitting in the $\left(
\mathbf{6},\mathbf{2}\right) $ of $USp(6)\otimes SU(2)_{H}$.

The resulting $\mathcal{N}=2$ $J_{3,M}^{\mathbb{R}}$ spectrum is composed by
$28$ $m=0$ real hypers' scalar degrees of freedom (rearranging in $7$
quaternionic hypermultiplet scalar), and by $n_{V}+1=2$ $m\neq 0$ and $%
n_{V}-1=0~m=0$ real vectors' scalar degrees of freedom, whose mass
degeneracy pattern thus confirms once again the Hessian splitting found in
\cite{TT} (no ``flat'' directions of non-BPS $Z\neq 0$ Hessian, implying
that the non-BPS $Z\neq 0$ critical points of $V_{BH,\mathcal{N}=2}$ in the
model $J_{3,M}^{\mathbb{R}}$ are \textit{all} stable).\medskip

For what concerns the other ``\textit{mirror}'' models, there is nothing
more to say. Indeed, $J_{3,M}^{\mathbb{C}}$ has $n_{V}=0$ and thus it
corresponds to a Reissner-N\"{o}rdstrom (extremal) BH with (graviphoton)
charges $p^{0}$ and $q_{0}$, only admitting $\frac{1}{2}$-BPS critical
points for $V_{BH,\mathcal{N}=2}$. Furthermore, as previously mentioned, $%
J_{3,M}^{\mathbb{H}}$ does not exist (\textit{at least} as far $d=4$ is
concerned), and $stu$ is \textit{self-mirror}: $stu_{,M}=stu$.

\section{\label{Conclusion}\textbf{Conclusion}}

In the present paper, in order to understand more in depth the nature of the
non-BPS solutions to attractor equations in $\mathcal{N}=8$, $d=4$
supergravity, we considered the supersymmetry reduction down to $\mathcal{N}%
=2$, $d=4$ magic supergravities (and their \textit{``mirror''} theories).
The multiplets' content is given by $n_{V}$ vector supermultiplets, whose
complex scalars span a special K\"{a}hler manifold of dimension $n_{V}$, and
by $n_{H}$ hypermultiplets, whose quaternionic scalars span a quaternionic
K\"{a}hler manifold of dimension $n_{H}$.

The mass spectrum of vector multiplets' scalars (the only relevant for the
Attractor Mechanism in ungauged supergravities) in $\mathcal{N}=2$ magic
supergravities has been studied in \cite{BFGM1}. By taking into account also
the ``hidden'' modes truncated away in the supersymmetry reduction $\mathcal{%
N}=8\longrightarrow \mathcal{N}=2$, the splittings of the $\mathcal{N}=2$
spectra should reproduce the splittings of the full spectra of the $70$ real
scalars of the parent $\mathcal{N}=8$ theory. We have shown how this works,
and in particular we reproduced the result of \cite{TT} about the mass
splitting of the modes of the $\mathcal{N}=2$ non-BPS $Z\neq 0$ Hessian.

By the supersymmetry reduction $\mathcal{N}=8\longrightarrow \mathcal{N}=2$,
the eventual instability of $\mathcal{N}=2$ non-BPS $Z\neq 0$ solutions to
attractor equations studied in \cite{TT} should reflect in a possible
instability of $\mathcal{N}=8$ non-BPS critical points of $V_{BH}$ in $%
\mathcal{N}=8$, $d=4$ supergravity.

On the other hand, by assuming that supersymmetry determines the $\mathcal{N}%
=8$, $\frac{1}{8}$-BPS critical points to be stable, it is possible to argue
that the $\mathcal{N}=2$ non-BPS $Z=0$ critical points of $V_{BH,\mathcal{N}%
=2}$ should be stable (beside the $\mathcal{N}=2$, $\frac{1}{2}$-BPS
critical points, whose stability is known after \cite{FGK}).
Correspondingly, when covariantly differentiating $V_{BH,\mathcal{N}=2}$
beyond the second order, the eventual ``flat'' directions of the non-BPS $%
Z=0 $ Hessian should suitably lift to directions with strictly positive
eigenvalues, or remain ``flat'' at all orders. Among the considered models,
only the $\mathcal{N}=2$, $d=4$ $stu$ supergravity (having $\left(
n_{V},n_{H}\right) =\left( 3,4\right) $, and thus \textit{self-mirror})
exhibit non-BPS, $Z=0$ critical points stable already at the Hessian level.
This can be understood by noticing that in such an $\mathcal{N}=2$ framework
\textit{triality symmetry} puts non-BPS $Z=0$ critical points on the very
same footing of $\frac{1}{2}$-BPS critical points, which are always stable
\cite{FGK} and thus do not have any ``flat'' direction at all.

We conclude by saying that our analysis could be applied to non-BPS critical
points of $V_{BH}$ in $2<\mathcal{N}<8$, ($d=4$) extended supergravities,
eventually comparing the $\mathcal{N}=8$ non-BPS spectrum with spectra
arising in $2<\mathcal{N}<8$ theories obtained by consistent supersymmetry
reductions (along the lines of \cite{ADF}), as done in \cite{ADF2} for the $%
\mathcal{N}=8$, $\frac{1}{8}$-BPS spectrum. Ultimately, such a procedure
could be performed for the $\mathcal{N}=1$, $d=4$ reduction of these
theories, especially of the $\mathcal{N}=2$ SK $d$-geometries \cite{ADFT-2}.

\section*{\textbf{Acknowledgments}}

The work of S.F. has been supported in part by European Community Human
Potential Program under contract MRTN-CT-2004-005104 \textit{``Constituents,
fundamental forces and symmetries of the universe''} and the contract
MRTN-CT-2004-503369 \textit{``The quest for unification: Theory Confronts
Experiments''}, in association with INFN Frascati National Laboratories and
by D.O.E. grant DE-FG03-91ER40662, Task C.

The work of A.M. has been supported by a Junior Grant of the \textit{%
``Enrico Fermi''} Center, Rome, in association with INFN Frascati National
Laboratories, and in part by D.O.E. grant DE-FG03-91ER40662, Task C.

A.M. would like to thank the Department of Physics and Astronomy, University
of California at Los Angeles, where this project was completed, for kind
hospitality and stimulating environment.

We would like also to acknowledge Restaurant \textit{``Lawry's-The Prime
Rib''} in Beverly Hills, for its inspiring atmosphere.

\end{document}